# The Absolutism of Space and Time with Variable Scalars and Gravitational Theory as well as Cosmology Established in Flat Reference Frame

## (2)

### Mei  Xiaochun

(Department of Physics, Fuzhou University, China, E-mail: mxc001@163.com)

## Content

   **Authors who are not interested in the concept analysis of space-time**
   **and gravitation can start from Section 4 after reading introduction**

# Section 4　The Revised Formulas of Newtonian Gravity Deduced from the Schwarzschild Solution of the Einstein's Equation of Gravitation fields

**1. Introduction**

　　In the current general relativity, the calculations of concrete problems are carried out in the curve space-time. Because the stander rulers and clocks can not be defined in the curve space-time, these results of calculations are meaningless actually in measurements. In order to comparing theoretical predictions calculated in the curve space-time with the experiments which are carried out on the earth reference frame which is nearly flat, we should transform the theoretical results into flat space-time to describe. A method is proposed to transform the geodesic equation described by the Schwarzschild solution of the Einstein's equation of gravitational field into flat reference frame to describe, and the revised formula of the Newtonian gravitation is obtained. By means of the formula, the experimental verifications of general relativity such as the perihelion precession of the Mercury, the deviation of light and the delay of radar wave in the gravitational field of the sun can also be explained well. The result indicates that photon is acted by repulsion force in gravitational field with positive potential. Light's speed in gravitational field is less than its speed in vacuum. In this theory, all singularities appearing in general relativity disappears completely and the black holes with singularities are considered not to exist. Thought there are still some black stars with great potential barriers of gravitation which photons can not surpass. A new formula of gravitational redshift and the revised formula of the Doppler shift in gravitational are obtained. An experiment is proposed to verify the change of light's speed in the gravitational field of the earth, which can be considered as the new verification for general relativity in the weak field with spherical symmetry.

**2. Impropriety of direct calculation and measurement in curved space-time**

　　According to the Einstein's theory of gravity, the space-time of gravitational field is curved. But can we determine the curvature of space-time through the direct measurements of observers who are located in gravitational fields? The answer is no. This because that we should define standard rulers and clocks for the meaningful measurements. But standard rulers and clocks can not be defined in curved space-time, it only be done in flat space-time. If there are no the definitions of standard rulers and clocks, we can not establish the meaningful concepts of distances, angles and time intervals and so on with measurement significance. However, when we move the standard rulers and clocks defined in flat space-time into gravitational fields, the rulers and clocks would become "curved" synchronously so that we can not use them to measure space-time's curvature of gravitational fields. The measurement's results in curved space-time are meaningless in practices if they can not be compared with that in flat space-time. On the other hand, according to the strong principle of equivalence, we may introduce local inertial reference frames in gravitational fields in which standard rulers and clocks can be defined. But in local inertial reference frames, gravitational fields are considered to disappear. So even though we may define standard rulers and clocks in local inertial reference frames, they are useless. In fact, in the current calculations of gravitational problems based on general relativity, no standard rulers and clocks of the local inertial reference frames are used. We always calculate the problems directly in curved space-time.

　　On the other hand, the gravitational field of the earth is quite weak so that it can be considered to be approximately flat. In fact, all experimental verifications about the Einstein's theory of gravitation are completed on the earth. So we should transform the theoretical predictions of general relativity calculated in curved space-time into that in flat space-time, then comparing the results with the experiments carried on the earth. Only in this way, we can say that the Einstein's theory of gravitation is correct or wrong. At present, all calculations of concrete problems in the Einstein's theory of gravitation are carried out in the curved space-time. For example, the angles of the perihelion precession of the Mercury and the deviation of light in the solar gravitational field, the time intervals of delay effect of radar waves and so on, all of theses quantities are defined in curved space-time. Unfortunately, these calculation results in curved space-time are always compared directly with the experiments carried out on the earth without transforming them into the results of flat space-time at present. Therefore, it is improper to affirm that the Einstein's theory of gravity has been verified before we do them. Otherwise the judgment would be relaxed and trustless. This is a principle problem, but it is neglected completely at present.

The Einstein's theory of gravity based on curved space-time has become the mainstream of gravitational research. But there exist same insurmountable difficulties in it, just as the problems of renormalization and singularity, the definition of gravitational field's energy and so on. On the other hand, because the Einstein's equations of gravitational fields are nonlinear, the idea to establish the theory of gravity based on flat space-time is always attractive. From the 1940's, lots of persons tried to re-establish the gravitational theory in flat space-time[14]. These theories are equivalent with the Einstein's theory under the conditions of weak fields, but in general situations they are different. But at present, no experiments can prove that these theories are superior to the Einstein's theory. So in light of common viewpoint, the space-time of gravitational fields should be described by non-Euclidean geometry. The flat space-time is always regarded as the boundary condition far away from gravitational fields.

On the other hand, according to the non-Euclidean geometry, the metrics of curved space-time can not be transformed into that of flat space-time in general, otherwise the curved space would not be the real curved one. This conclusion seems to indicate that gravitational fields can not be described in flat space-time. However, this impossibility only means that we can not transform the whole metrics of gravitational fields from the non-Euclidean into the Euclidean. But we can always transform a curve described in curved space into that described in flat space. In fact, only by observing the object's motions in gravitational fields, we can comprehend the nature of gravitational field's space-time. It is unnecessary for us to transform whole curved space-time into flat space-time. According to the general theory of relativity, objects always move along the geodetic lines in gravitational fields. As long as we transform the geodetic lines defined in curved space-time into the curved lines or the motion equations defined in flat space-time, we can transform the theory of gravity described in curved space-time into that described in flat space-time. The most important is that in this way, all singularities and morbid characters in general relativity disappears completely and the theory of gravitation would become a health theory.

## 2. The Schwarzschild solution described in flat space-time

Now let's discuss how to transform the Schwarzschild metric of the Einstein's equation of gravitational field into flat space-time to description. According to the general theory of relativity, the Schwarzschild metric of static mass distribution with spherical symmetry (external solution) is

$$ds^2 = c^2\left(1 - \frac{\alpha}{r}\right)dt^2 - \left(1 - \frac{\alpha}{r}\right)^{-1} dr^2 - r^2\left(\sin^2\theta d\varphi^2 + d\theta^2\right) \qquad (4.1)$$

Here $\alpha = 2GM/c^2$. According to the familiar results in general relativity, let $\theta = \pi/2$ and substitute Eq.(4.1) into the equation of geodetic line, we get the integrals

$$c\left(1-\frac{\alpha}{r}\right)\frac{dt}{ds}=\varepsilon \qquad r^2\frac{d\varphi}{ds}=\frac{L}{c} \qquad (4.2)$$

Here $\varepsilon$ and $L$ are constants. From above two formulas, we can eliminate linear element $ds$ and get

$$r^2\left(1-\frac{\alpha}{r}\right)^{-1}\frac{d\varphi}{dt}=\frac{L}{\varepsilon} \qquad (4.3)$$

Defining

$$d\tau=\left(1-\frac{\alpha}{r}\right)dt \qquad (4.4)$$

Considering $\tau$ as the proper time, $t$ as the coordinate time and let $\varepsilon=1$, we have from Eq.(4.2)

$$ds=cd\tau \qquad (4.5)$$

Thus, Eq.(4.4) becomes

$$r^2\frac{d\varphi}{d\tau}=L \qquad (4.6)$$

Here $L$ is the angel momentum of unit mass. Eq.(4.6) is just the conservation formula of angel momentum. Let's first discuss the motions of particles with static masses. By using Eq.(4.6), Eq.(4.1) can be written as

$$\left(1-\frac{\alpha}{r}\right)\left(\frac{dt}{d\tau}\right)^2-\left(1-\frac{\alpha}{r}\right)^{-1}\left(\frac{dr}{cd\tau}\right)^2-r^2\left(\frac{d\varphi}{cd\tau}\right)^2=1 \qquad (4.7)$$

From Eq.(4.5) and (4.7), we get

$$\left(\frac{dr}{d\tau}\right)^2=\frac{c^2\alpha}{r}\left(1-\frac{L^2}{\alpha c^2 r}+\frac{L^2}{c^2 r^2}\right) \qquad (4.8)$$

By the differential about $d\tau$ in the formula above, we get

$$\frac{d^2r}{d\tau^2}-\frac{L^2}{r^3}=-\frac{c^2\alpha}{2r^2}\left(1+\frac{3L^2}{c^2 r^2}\right) \qquad (4.9)$$

It should be noted that all quantities in Eq.(4.9) are defined in the caved space-time. In order to describe the geodetic line equation in flat space-time, coordinate transformation is needed. Let $r_0$, $\varphi_0$ and $t_0$ represent the space-time coordinates of flat space-time, due to the invariability of the 4-diamention interval $ds^2$, we have

$$ds^2=c^2dt_0^2-dr_0^2-r_0^2 d\varphi_0^2=c^2\left(1-\frac{\alpha}{r}\right)dt^2-\left(1-\frac{\alpha}{r}\right)^{-1}dr^2-r^2 d\varphi^2 \qquad (4.10)$$

It can be seen that the forms of the third items on the two sides of formula above are completely the same. The difference is only on the symbols. So we can take $r_0=r$, $\varphi_0=\varphi$ and then get the transformation between times $t_0$ and $t$

$$c^2 dt_0^2=c^2\left(1-\frac{\alpha}{r}\right)dt^2+\left[1-\left(1-\frac{\alpha}{r}\right)^{-1}\right]dr^2 \qquad (4.11)$$

On the other hand, by considering Eq.(4.4) and (4.8), we have

$$dr = c\left(1-\frac{\alpha}{r}\right)\sqrt{\frac{\alpha}{r}\left(1-\frac{L^2}{\alpha c^2 r}+\frac{L^2}{c^2 r^2}\right)}\,dt \qquad (4.12)$$

Put the formula into Eq.(4.11), we have

$$dt_0 = \sqrt{\left(1-\frac{\alpha}{r}\right)\left[1-\frac{\alpha^2}{r^2}\left(1-\frac{L^2}{\alpha c^2 r}+\frac{L^2}{c^2 r^2}\right)\right]}\,dt \qquad (4.13)$$

Comparing it with Eq.(4.4), we get

$$d\tau = \left(1-\frac{\alpha}{r}\right)^{\frac{1}{2}}\left[1-\frac{\alpha^2}{r^2}\left(1-\frac{L^2}{\alpha c^2 r}+\frac{L^2}{c^2 r^2}\right)\right]^{-\frac{1}{2}}\,dt_0 \qquad (4.14)$$

Because we have defined $r_0 = r$, all quantities on the light side of formula above have been defined in flat space-time. On the other hand, in the classical Newtonian theory of gravity, the motion equations of unit mass in the plane polar coordinates are

$$\frac{d^2 r}{dt^2} - r\left(\frac{d\varphi}{dt}\right)^2 = -\frac{c^2\alpha}{2r^2} \qquad (4.15)$$

$$\frac{1}{r}\frac{d}{dt}\left(r^2\frac{d\varphi}{dt}\right) = 0 \qquad \text{or} \qquad r^2\frac{d\varphi}{dt} = L \qquad (4.16)$$

Substituting Eq.(4.16) into Eq.(4.15), we obtain

$$\frac{d^2 r}{dt^2} - \frac{L^2}{r^3} = -\frac{c^2\alpha}{2r^2} \qquad (4.17)$$

Comparing Eq.(4.17) with Eq.(4.9), we know that besides the revised item on the light side of Eq.(4.9), as long as take $\tau \leftrightarrow t$, the forms of Eqs.(4.9) and (4.17) are completely the same. So we can write Eq.(4.9) in the vector form similar to the Newtonian gravitational theory

$$\frac{d^2\vec{r}}{d\tau^2} = -GM\left(1+\frac{3L^2}{c^2 r^2}\right)\frac{\vec{r}}{r^3} \qquad (4.18)$$

In the formula, all quantities in the formula have been defined in flat space-time. Let $u = 1/r$ and by considering Eq.(4.6), the formula above can be transformed into

$$\frac{d^2 u}{d\varphi^2} + u = \frac{GM}{L^2}\left(1+\frac{3L^2 u^2}{c^2}\right) \qquad (4.19)$$

This formula can be used to describe the perihelion precession of the Mercury. Now let's prove that the effect of special relativity has been taken into account in Eq.(4.18). From Eqs.( 4.8), (4.10)----(4.14), we can obtain

$$V_r^2 = \left(\frac{dr}{dt_0}\right)^2 = \left(\frac{dr}{d\tau}\frac{d\tau}{dt_0}\right)^2 = \frac{c^2\alpha}{r}\left(1-\frac{\alpha}{r}\right)\left(1-\frac{L^2}{\alpha c^2 r}+\frac{L^2}{c^2 r^2}\right)\left[1-\frac{\alpha^2}{r^2}\left(1-\frac{L^2}{\alpha c^2 r}+\frac{L^2}{c^2 r^2}\right)\right]^{-1} \qquad (4.20)$$

$$V_\varphi^2 = \left(r\frac{d\varphi}{dt_0}\right)^2 = \left(r\frac{d\varphi}{d\tau}\frac{d\tau}{dt_0}\right)^2 = \frac{L^2}{r^2}\left(1-\frac{\alpha}{r}\right)\left[1-\frac{\alpha^2}{r^2}\left(1-\frac{L^2}{\alpha c^2 r}+\frac{L^2}{c^2 r^2}\right)\right]^{-1} \quad (4.21)$$

$$V^2 = V_r^2 + V_\varphi^2 = \frac{c^2\alpha}{r}\left(1-\frac{\alpha}{r}\right)\left(1+\frac{L^2}{c^2 r^2}\right)\left[1-\frac{\alpha^2}{r^2}\left(1-\frac{L^2}{\alpha c^2 r}+\frac{L^2}{c^2 r^2}\right)\right]^{-1} \quad (4.22)$$

$$1-\frac{V^2}{c^2} = \left(1-\frac{\alpha}{r}\right)\left[1-\frac{\alpha^2}{r^2}\left(1-\frac{L^2}{\alpha c^2 r}+\frac{L^2}{c^2 r^2}\right)\right]^{-1} \quad (4.23)$$

Comparing with Eq.(4.14), we get

$$d\tau = \sqrt{1-\frac{V^2}{c^2}}dt_0 \quad (4.24)$$

This is just the formula of time delay in special relativity. At last, let $t_0 \to t$, Eq.(4.18) can be written as

$$\frac{d\bar{p}}{dt} = -GMm_0\left(1+\frac{3L^2}{c^2 r^2}\right)\sqrt{1-\frac{V^2}{c^2}}\frac{\vec{r}}{r^3} \quad (4.25)$$

Here $m_0$ the static mass of a particle. The formula can be regarded as the revision of the Newtonian formula of gravity. In this way, Eq.(4.6) can be written as

$$r^2\frac{d\varphi}{d\tau} = \frac{r^2\dot{\varphi}}{\sqrt{1-V^2/c^2}} = L \quad (4.26)$$

So in the center gravitational fields, the classical angle momentum $m_0 r^2 \dot{\varphi}$ is not a constant again. It should be divided by a contraction factor of special relativity.

### 3. The motion of particle in gravitational field with spherical symmetry

The problem of energy conservation is discussed below. For simplification, we only discuss the situation that a particle moves along the radius vector direction with $L=0$. By considering Eq.(4.23) in this case, Eq.(4.25) becomes

$$\frac{d\bar{p}}{dt} = -\frac{GMm_0}{\sqrt{1+\alpha/r}}\frac{\vec{r}}{r^3} \quad (4.27)$$

By producing $d\bar{r}$ on the two sides of Eq.(4.27), the potential energy of gravitational field is

$$U(r) = -\int \vec{F}\cdot d\bar{r} = \int \frac{m_0 c^2 \alpha}{2\sqrt{1+\alpha/r}}\frac{\vec{r}}{r^3}\cdot d\bar{r} = -m_0 c^2\sqrt{1+\frac{\alpha}{r}} + A_1 \quad (4.28)$$

Here $A_1$ is a constant. When $r \to \infty$, we have $U \to 0$ and get $A_1 = m_0 c^2$. The integral on the left side of Eq.(4.27) can be written as

$$T = \int \frac{d\bar{p}}{dt}\cdot d\bar{r} = \int \frac{d\bar{p}}{dt}\cdot\frac{d\bar{r}}{dt}dt = \int \vec{V}\cdot d\bar{p} = \vec{V}\cdot\bar{p} - \int \bar{p}\cdot d\vec{V}$$

$$= \frac{m_0 V^2}{\sqrt{1-V^2/c^2}} + m_0 c^2\sqrt{1-\frac{V^2}{c^2}} + A_2 \quad (4.29)$$

Here $T$ is just the dynamic energy of particle. When $r \to \infty$, we have $V=0$ and get $A_2 = -m_0 c^2$.

So the law of energy conservation of a particle in the gravitational field can be written as

$$T + U = \frac{m_0 V^2}{\sqrt{1 - V^2/c^2}} + m_0 c^2 \left( \sqrt{1 - \frac{V^2}{c^2}} - 1 \right) - m_0 c^2 \left( \sqrt{1 + \frac{\alpha}{r}} - 1 \right) = 0 \qquad (4.30)$$

When $\alpha/r \ll 1$, $V \ll c$, from Eq.(4.30) we get the classical law of energy conservation in the Newtonian theory of gravity

$$\frac{m_0 V^2}{2} - \frac{GM m_0}{r} = 0 \qquad (4.31)$$

For the situation with $L \neq 0$, we can also calculate the problem in the weak field with $\alpha/r \ll 1$. By remaining items with the orders up to $r^{-2}$, we have

$$U(r) = -\int \vec{F} \cdot d\vec{r} = \int \frac{m_0 c^2 \alpha}{2\sqrt{1 + \alpha/r}} \left( 1 + \frac{3L^2}{c^2 r^2} \right) \frac{\vec{r}}{r^3} \cdot d\vec{r}$$

$$= -m_0 c^2 \left\{ \sqrt{1 + \frac{\alpha}{r}} - \frac{3L^2}{c^2 \alpha^2} \left( 1 + \frac{\alpha}{r} \right) \left[ \frac{1}{6}\left(1 + \frac{\alpha}{r}\right)^2 - \frac{2}{3}\left(1 + \frac{\alpha}{r}\right)^{1/2} + 1 \right] \right\} + A_1 \qquad (4.32)$$

So the law of energy conservation is

$$E = \frac{m_0 V^2}{\sqrt{1 - V^2/c^2}} + m_0 c^2 \left\{ \sqrt{1 - \frac{V^2}{c^2}} - \sqrt{1 + \frac{\alpha}{r}} \right.$$

$$\left. + \frac{3L^2}{c^2 \alpha^2} \left( 1 + \frac{\alpha}{r} \right) \left[ \frac{1}{6}\left(1 + \frac{\alpha}{r}\right)^2 - \frac{2}{3}\left(1 + \frac{\alpha}{r}\right)^{1/2} + 1 \right] \right\} \qquad (4.33)$$

Here $E$ is a constant.

Let's now discuss the motion of an experimental particle along the direction of radius in the curved Schwarzschild coordinate. After that, we discuss the problem in flat space-time. From Eq.(4.4) and (4.8) in curved space-time, when $L = 0$, we get particle's speed

$$V = \frac{dr}{dt} = \pm c \sqrt{\frac{\alpha}{r}} \left( 1 - \frac{\alpha}{r} \right) \qquad (4.34)$$

Within the region $r > \alpha$, when the particle moves along the positive direction of radius vector, the formula takes positive sign. When the particle moves along the negative direction of radius vector, the formula takes negative sign. Within the region $r < \alpha$, when the particle moves along the positive direction of radius vector, the formula takes negative sign and when the particle moves along the negative direction of radius vector, the formula takes positive sign. It is obvious that when $r \to \infty$, we have $V = 0$. When $r = \alpha$, i.e., the particle reaches the event horizon, we also have $V = 0$. Within the region $\sqrt{\alpha/r}(1 - \alpha/r) < 1$, particle's speed is less than light's speed in vacuum. Within the region $\sqrt{\alpha/r}(1 - \alpha/r) > 1$, the particle's speed surpasses light's speed in vacuum. When $r \to 0$, we have $V \to \infty$. From the formula above, particle's acceleration is

$$a = \frac{dV}{dt} = -\frac{1}{2} \frac{c^2 \alpha}{r^2} \left( 1 - \frac{\alpha}{r} \right) \left( 1 - \frac{3\alpha}{r} \right) \qquad (4.35)$$

At the points $r = 3\alpha$ and $r = \alpha$, particle's accelerations are zero. Within the region $r > 3\alpha$, we have $a < 0$. It means that particle is acted by gravitation so its speed is increased. Within the region $\alpha < r < 3\alpha$, we have $a > 0$. The particle is acted by repulsion force and its speed is decreased. (It is inconceivable in this case that gravitational force becomes repulsion force.). When $r = \alpha$, we have $V = a = 0$. In this case, particle is at rest on the event horizon without the action of force. When $r < \alpha$, we have $a < 0$, particle is acted by gravitation and moves towards the center of gravitational field. When $r \to 0$, we also have $a \to \infty$.

Let's consider the integral of Eq.(4.35). Because there exist singularity at point $r = \alpha$, the integrals should be taken individually in different regions. Within the region $r > \alpha$ with the initial condition $r = r_0$ when $t = 0$, the integral of Eq.(4.34) is

$$ct = \pm \frac{1}{\sqrt{\alpha}} \left[ \frac{2}{3}\left(r^{3/2} - r_0^{3/2}\right) + 2\alpha\left(\sqrt{r} - \sqrt{r_0}\right) + \alpha^{3/2} \ln \frac{(\sqrt{r} - \sqrt{\alpha})(\sqrt{r_0} + \sqrt{\alpha})}{(\sqrt{r} + \sqrt{\alpha})(\sqrt{r_0} - \sqrt{\alpha})} \right] \qquad (4.36)$$

Suppose the particle falls down in gravitational field, we take negative sign in the formula above. When $r \to \alpha$, we heve $t \to \infty$. It means that the particle needs an infinite time to reach the event horizon. Then suppose that the particle's initial position is at $r_0 \to \alpha$. We take positive sign so that the particle moves up apart from the event horizon. Suppose that the particle reaches $r = \alpha + \Delta$ point at a certain time, $\Delta$ may be a very small but limited value. Because the third item of (4.36) would become infinite when $r_0 \to \alpha$, we have $t \to \infty$ in this case. It means that the particle on the event horizon can not move up apart from the event horizon actually.

Then we discuss particle's motion within the event horizon with $r < \alpha$. Suppose that the particle is at point $r = r_0$ when $t = 0$, the integral of Eq.(4.34) is

$$ct = \pm \frac{1}{\sqrt{\alpha}} \left[ \frac{2}{3}\left(r^{3/2} - r_0^{3/2}\right) + 2\alpha\left(\sqrt{r} - \sqrt{r_0}\right) + \alpha^{3/2} \ln \frac{(\sqrt{\alpha} - \sqrt{r})(\sqrt{\alpha} + \sqrt{r_0})}{(\sqrt{\alpha} + \sqrt{r})(\sqrt{\alpha} - \sqrt{r_0})} \right] \qquad (4.37)$$

We take $r_0 = 0$ at first. According to Eq.(4.34), the initial speed of particle would be infinite. Suppose that particle moves up, we take negative sign in the formula above and get

$$ct = -\frac{1}{\sqrt{\alpha}} \left( \frac{2}{3} r^{3/2} + 2\alpha \sqrt{r} + \alpha^{3/2} \ln \frac{\sqrt{\alpha} - \sqrt{r}}{\sqrt{\alpha} + \sqrt{r}} \right) \qquad (4.38)$$

It is easy to verify that the time is a positive number within the region $r < \alpha$ in the formula above. When particle reaches the event horizon, we have $t \to \infty$. Then suppose that the initial position of particle is on the event horizon with $r_0 = \alpha$. By the action of gravitation, particle moves down towards the center of gravitational field. We take positive sign in Eq.(4.37). Suppose that the particle reaches $r = \alpha - \Delta$ point at a certain time, $\Delta$ may be a very small but limited value. Because the third item of Eq.(4.37) would become infinite when $r_0 \to \alpha$, we have $t \to \infty$ in this case. It means that the particle on the event horizon can not move down continuously and then collapse at the center singularity of gravitational field after it reaches the event horizon. The results indicate that the event horizon is actually an attractive plane for moving particles. The particles could not leave the event horizon along both up and down directions after they had reached the event horizon. But this result is neglected in the current theory of black holes. According to the current understanding, any particles would move towards to the renter of gravitational field so that they would collapse at the center singularity at last.

It is obvious that there are some things irrational in the processes of particle's motions in the gravitational fields, besides the singularity of the event horizon. For example, some time the particle would be accelerated and some time it would be decelerated outside the event horizon. Especially particles would move in the speeds surpassing light's speed in vacuum, even move in an infinite speed. In the current theory, those problems are attributed to the improper selections of coordinates. In order to eliminate those defects, some coordinate systems just as the Eddington's and the Kruskal's coordinate systems[15] are introduced. In new coordinate systems, though the singularities on the event horizons may be eliminated, they can not yet be eliminated at the original point $r = 0$. Hawking etc. even proved that it was impossible to eliminate all singularities in the general theory of relativity[16].

However, as shown in former paper, it should be emphasized again that the arbitrary transformation in four dimension space-time is impossible for gravitational problems. This kind of transformations would introduce arbitrary inertial forces, and the inertial forces are considered to equivalent to arbitrary gravitational fields. So after the transformations, the new gravitational fields are not equal to original ones. In the current theories of black holes, in order to eliminate singularity on the event horizon, the freely falling Novikow or Lemaitre coordinates are introduced. Because there are no the event horizons, observers can enter the event horizons without any felling. After that, according to the current understanding, they would be attracted into the center of black holes and be torn into the pieces by so-called tide forces at last. But in the Schwarzschild coordinate, the observers would stop on the event horizon forever. However, the observer's life and death are absolute events, what are observer's fates?  In principle, we can find infinite coordinate systems in which the singularities on the event horizons can be eliminated. But none of theses metrics are with spherical symmetry except the Schwarzschild coordinate. How can they represent the gravitational field with spherical symmetry? On the other hand, in the curved coordinate systems, no matter in the Novikow or Lemaitrewe coordinate systems, we can not define stander rulers and clocks. In theses curved coordinate systems, the speeds of clocks located at different places are not the same so that the measurements of time intervals are meaningless. For example, we say that an observer freely falling down the gravitational field would spend infinite time to reach the event horizon. Because the clock's speed in the moving reference frame changes continuously, what is the real meaning of infinite time?  Because only in the flat space-time we can define stander rulers and clocks, only in the flat reference frames outside the gravitational fields, the calculations and measurements for objects moving in the gravitational fields are meaningful. So the problems of gravitation should be transformed into the flat space-time for discussion.

Now let's discuss the motion of a particle in the gravitational field from the angle of observers in flat space-time outside gravitational field. Suppose that a particle falls freely along the radium direction of gravitational field, its velocity and acceleration are individually

$$V = \frac{dr}{dt} = -c\sqrt{\frac{\alpha}{r}}\left(1 + \frac{\alpha}{r}\right)^{-1/2} \qquad (4.39)$$

$$a = -\frac{1}{2}\frac{c^2\alpha}{r^2}\left(1 + \frac{\alpha}{r}\right)^{-2} \qquad (4.40)$$

It is known that when $r \to \infty$, we have $V = 0$ and $a = 0$. Suppose when $t = 0$ the particle is at point $r = r_0$, by taking the integral of Eq.(4.39), we get

$$ct = \frac{2}{3\sqrt{\alpha}}\left[(r_0 + \alpha)^{3/2} - (r + \alpha)^{3/2}\right] \qquad (4.41)$$

It is obvious that every thing is normal within the region $r > 0$. The particle is monotonously accelerated by gravitation. There is no any singularity in the whole space-time. When particle arrives at the original point $r = 0$, we have

$$V = -\lim_{x \to \infty} \frac{c\sqrt{\alpha/r}}{\sqrt{1+\alpha/r}} \to -c \qquad a = -\lim_{x \to \infty} \frac{c^2 x^2}{2\alpha(x+1)^2} \to -\frac{c^2}{2\alpha} \qquad (4.42)$$

$$F = -\lim_{x \to \infty} \frac{c^2 x^2}{2\alpha(x+1)} \to -\lim_{x \to \infty} \frac{c^2 x}{2\alpha} \to -\infty \qquad (4.43)$$

It shows that the speed of particle tends to light's speed in vacuum. Acceleration is also finite. So within the region $0 < r \leq \infty$, the motion of particles with static masses are continuous. Only at point $r = 0$, the force acted on particles are infinite. But this infinite also appears in the Newtonian theory of gravitation, having nothing to do with space-time singularity. When a particle moves along the positive direction of radius vector, as long as its velocity satisfies Eq.(4.39), the particle would escape gravitational field and has a speed $V = 0$ when it reach the place $r \to \infty$. That is to say, after the Schwarzschild solution is transformed into flat space-time to describe, for the motion of particles with static mass, all space-time singularities disappear. So it is obvious that singularities appearing in the Einstein's theory of gravity are actually caused by the descriptive method in curved space-time. As long as we describe the problems og gravitation in flat space-time, all singularities are canceled. The gravitational field itself has no singularities. In real and physical world, singularity is not allowed to exist.

### 4. Photon's motions in gravitational field with spherical symmetry

The motion equation of photon in flat space-time is discussed as follows. For photons, we have $ds = 0$. In this case, we take $d\tau$ as parameter to describe the equation of geodetic line. By solving the Einstein's equation of gravity with spherical symmetry, we obtain

$$\left(1 - \frac{\alpha}{r}\right)\frac{dt}{d\tau} = \varepsilon \qquad r^2 \frac{d\varphi}{d\tau} = L \qquad (4.44)$$

Let $\varepsilon = 1$ similarly, we have

$$d\tau = \left(1 - \frac{\alpha}{r}\right) dt \qquad (4.45)$$

Because of

$$ds^2 = c^2\left(1 - \frac{\alpha}{r}\right) dt^2 - \left(1 - \frac{\alpha}{r}\right)^{-1} dr^2 - r^2 d\varphi^2 = 0 \qquad (4.46)$$

the formula above can be written as

$$c^2\left(1 - \frac{\alpha}{r}\right)\left(\frac{dt}{d\tau}\right)^2 - \left(1 - \frac{\alpha}{r}\right)^{-1}\left(\frac{dr}{d\tau}\right)^2 - \left(r\frac{d\varphi}{d\tau}\right)^2 = 0 \qquad (4.47)$$

From the formula, we obtain

$$\left(\frac{dr}{d\tau}\right)^2 = c^2\left[1 - \left(1 - \frac{\alpha}{r}\right)\frac{L^2}{c^2 r^2}\right] \qquad (4.48)$$

By taking the differential of Eq.(4.48) about $d\tau$, we get

$$\frac{d^2 r}{d\tau^2} - \frac{L^2}{r^3} = -\frac{3\alpha L^2}{2r^4} \tag{4.49}$$

$\vec{V}_c$  4.45  4.48

$$V_{rc}^2 = \left(\frac{dr}{dt}\right)^2 = c^2\left(1-\frac{\alpha}{r}\right)^2\left[1-\left(1-\frac{\alpha}{r}\right)\frac{L^2}{c^2 r^2}\right] \tag{4.50}$$

$$V_{\varphi c}^2 = \left(r\frac{d\varphi}{dt}\right)^2 = \left(r\frac{d\varphi}{d\tau}\frac{d\tau}{dt}\right)^2 = \left(1-\frac{\alpha}{r}\right)^2\frac{L^2}{r^2} \tag{4.51}$$

$$V_c = \pm\sqrt{\left(\frac{dr}{dt}\right)^2 + \left(r\frac{d\varphi}{dt}\right)^2} = \pm c\left(1-\frac{\alpha}{r}\right)\sqrt{1+\frac{\alpha L^2}{c^2 r^3}} \tag{4.52}$$

It is obvious that the speed of light in the gravitational field would change with $V_c \neq c$ in general. Then suppose that photon moves along the radius vector's direction with $L=0$. The velocity of photon is

$$V = \frac{dr}{dt} = \pm c\left(1-\frac{\alpha}{r}\right) \tag{4.53}$$

When $r \to \infty$ we have $V = 0$. When $r = \alpha$ we also have $V = 0$, i.e., particle's speed would be zero when it reaches the event horizon. Within the region $\alpha/r > 2$, we have $V > c$. When $r \to 0$ we have $V \to \infty$, i.e., particle's speed becomes infinite when particle arrive at the center point of gravitational field. From the formula above, we can obtain particle's acceleration

$$a = \frac{dV}{dt} = \frac{c^2\alpha}{r^2}\left(1-\frac{\alpha}{r}\right) \tag{4.54}$$

So within the region $r > \alpha$, particle's acceleration is positive with $a > 0$. It indicates that particle is acted by repulsion force. When a particle falls down the gravitational field freely, its speed is decreased. When it moves up apart from the field, its speed is increased. Similarly, space-time has singularity at point $r = \alpha$, so the integral should also be carried out in the different regions. When particle falls down the gravitational field, the negative sign is taken in Eq.(4.53). Within the region $r > \alpha$, let $r = r_0$ when $t = 0$, the integral of Eq.(4.53) is

$$ct = r_0 - r - \alpha \ln\frac{r-\alpha}{r_0-\alpha} \tag{4.55}$$

It is known from Eq.(4.53) and (4.54) that photon's speed and acceleration are zero at point $r = \alpha$, so photons would stop at the event horizon. But from Eq.(4.54), it needs a infinite time for a photon to reach the event horizon. In this case, photons can not yet leave the event horizon, i.e., there exist so-called black holes. Within the event horizon with $r < \alpha$, particle's acceleration $a < 0$, so photon is acted by gravitation and its speed is increased. Suppose we have $r_0 \to \alpha$ when $t = 0$ and photon moves towards the center of gravitational field. The integral of Eq.(4.53) is

$$ct = r - r_0 + \alpha \ln\frac{\alpha-r}{\alpha-r_0} \tag{4.56}$$

Suppose the particle reaches $r = \alpha - \Delta$ point at a certain time, $\Delta$ may be a very small but limited value. Because the second item of Eq.(4.56) would become infinite when $r_0 \to \alpha$, we have $t \to \infty$ in this case. It means that the particle on the event horizon can not move up apart from the event horizon. So the particles could not leave the event horizon along both up and down directions after they had reached the event horizon. The event horizon is also an attractive plane for moving photons.

Then let's discuss how to describe photon's motion in flat space-time. For photon's motion, the metric in the flat space-time can be written as

$$ds^2 = c^2 dt_0^2 - dr_0^2 - r_0^2 d\varphi_0^2 = 0 \tag{4.57}$$

From Eqs.(4.46), we have

$$V_0^2 dt_0^2 - dr_0^2 - r_0^2 d\varphi = c^2\left(1 - \frac{\alpha}{r}\right)dt^2 - \left(1 - \frac{\alpha}{r}\right)^{-1} dr^2 - r^2 d\varphi^2 \tag{4.58}$$

Similarly, the forms of the third items on the two sides of above formula are the same completely, we can also let $r_0 = r$, $\varphi_0 = \varphi$ and get

$$c^2 dt_0^2 = c^2\left(1 - \frac{\alpha}{r}\right)dt^2 + \left[1 - \left(1 - \frac{\alpha}{r}\right)^{-1}\right]dr^2 \tag{4.59}$$

By using Eq.(4.50), we obtain

$$c^2 dt_0^2 = c^2\left(1 - \frac{\alpha}{r}\right)^2\left(1 + \frac{\alpha L^2}{c^2 r^3}\right)dt^2 \tag{4.60}$$

or

$$dt_0 = \left(1 - \frac{\alpha}{r}\right)\sqrt{1 + \frac{\alpha L^2}{c^2 r^3}}\, dt \qquad d\tau = \frac{dt_0}{\sqrt{1 + \alpha L^2 /(c^2 r^3)}} \tag{4.61}$$

Similarly, the motion equation of photons in the gravitational field with spherical symmetry can be written in the form of vector in flat space-time

$$\frac{d^2 \vec{r}}{d\tau^2} = -\frac{3\alpha L^2 \vec{r}}{2r^5} \tag{4.62}$$

Let $u = 1/r$, the formula above can be transformed into

$$\frac{d^2 u}{d\varphi^2} + u = \frac{3GM}{c^2} u^2 \tag{4.63}$$

The formula can be used to describe the deviation of light in the solar gravitational field. On the other hand, according to Eq.(4.50) and by taking approximation in weak field, we have

$$cdt = \frac{(1 + \alpha/r)[1 - \alpha L^2 /(c^2 r^3)]}{\sqrt{1 - L^2 /(c^2 r^2)}} dr \tag{4.64}$$

From the formula, the delay experiment of radar wave can be explained well[3]. But if using Eqs.(4.58) and (4.59) and taking approximation of weak field, we have (let $t_0 \to t$ in the formula so that $r$ and $t$ have been the coordinates of flat space-time.)

$$cdt = \frac{\sqrt{1+\alpha L^2/(c^2 r^3)}}{\sqrt{1-L^2/(c^2 r^2)+\alpha L^2/(c^2 r^3)}} dr \approx \frac{1-\alpha L^4/(4c^4 r^5)}{\sqrt{1-L^2/(c^2 r^2)}} dr \qquad (4.65)$$

It is obvious that if the formula above is used to calculate, the correct conclusion can not be reached. This result means that when the geodesic equation of photon in curve space-time is transformed into flat space-time, we can not describe correctly the delay effect of radar wave in the sun gravitational field. In order to describe photon's motion in flat space-time, we have to introduce same new hypotheses. At first, we suppose that the motion of photons is still described by formula (4.49) or (4.62), but the coordinates in the formula is defined in flat space-time. Meanwhile, in formula (4.49), we define proper time $d\tau$ as

$$d\tau = \frac{dt}{1+\alpha/r} \qquad (4.66)$$

Comparing with (4.45), it is known that when $\alpha/r \ll 1$, both are the same. So in flat space-time, we can describe explain both the deviation of light and the delay effect of radar wave in the weak gravitational field of the sun well by connecting the motion equation (4.62) and the definition 4.66 . But in strong fields, they may be completely different. Let's now prove that in light of (4.66), there is no any singularity for photon's motion in gravitational field again. Because we have supposed that all quantities in the formulas (4.44), (4.48) and (4.66) are defined in flat reference frame, photon's speed and acceleration in the gravitational field with spherical symmetry are

$$V_{rc}^2 = \left(\frac{dr}{dt}\right)^2 = \left(\frac{dr}{d\tau}\frac{d\tau}{dt}\right)^2 = \frac{c^2}{(1+\alpha/r)^2}\left[1-\left(1-\frac{\alpha}{r}\right)\frac{L^2}{c^2 r^2}\right] \qquad (4.67)$$

$$V_{\varphi c}^2 = \left(r\frac{d\varphi}{dt}\right)^2 = \left(r\frac{d\varphi}{d\tau}\frac{d\tau}{dt}\right)^2 = \frac{1}{(1+\alpha/r)^2}\frac{L^2}{r^2} \qquad (4.68)$$

$$V_c = \pm\sqrt{\left(\frac{dr}{dt}\right)^2+\left(r\frac{d\varphi}{dt}\right)^2} = \pm\frac{c}{1+\alpha/r}\sqrt{1+\frac{\alpha L^2}{c^2 r^3}} \qquad (4.69)$$

When $\alpha/r \ll 1$, the formulas above are the same as (4.50)~(4.52). When $L=0$, we have

$$V_c = \frac{dr}{dt} = \frac{dr}{d\tau}\frac{d\tau}{dt} = \pm\frac{c}{1+\alpha/r} \qquad (4.70)$$

$$a_c = \frac{dV}{dt} = \frac{dV}{dr}\frac{dV}{dt} = \frac{c^2\alpha}{(1+\alpha/r)^3 r^2} \qquad (4.71)$$

It can be seen that photon is acted by repulsion force and its speed is decreased in gravitational field. When photon arrives at the point $r=0$, we have $V_c = 0$. Its acceleration is also finite with

$$a_c = \lim_{x\to\infty}\frac{c^2 x^2}{\alpha(1+x)^3} \to 0 \qquad (4.72)$$

Suppose $r = r_0$ when $t = t_0 = 0$, we have

$$ct = r_0 - r + \alpha \ln\frac{r_0}{r} \qquad (4.73)$$

An infinite time is needed for photon to reach the point $r=0$. It means actually that photon can not be static actually. Similar to particles with statistic masses, all singularities of photon's motions disappear.

On the other hand, it should be noted that the result of light's speed being changed in gravitational field is deduced directly from the Einstein's equation of gravitation. Unfortunately this results is neglected at present so that same serious aftermaths are caused in gravitation and astrophysics. These problems will be discussed later.

## 5. The dynamic equation of photon in gravitational field with spherical symmetry

It is noted that 4.62 is not the motion equation of photon in gravitational field. According to the formula, we can write photon's momentum and force as formally

$$\vec{p} = m_0 \vec{V}_c = \frac{m_0 \vec{c}}{1 + \alpha/r} \qquad \vec{F} = \frac{d\vec{p}}{dt_0} = -\frac{1}{1 + \alpha/r} \frac{3\alpha L^2 \vec{r}}{2r^5} \qquad (4.74)$$

When photon moves along the radius vector direction with $L = 0$, we have $\vec{F} = 0$. It seems that photon is not acted by force. But according to Eq.(4.71), photon has acceleration, so it should be acted by force in this case. Therefore, Eq.(4.62) is not the dynamic equation of photon. On the other hand, by considering (4.66), we can write Eq.(4.62) as

$$\frac{d^2 \vec{r}}{dt^2} = -\frac{1}{(1 + \alpha/r)^2} \frac{3\alpha L^2 \vec{r}}{2r^5} + \frac{1}{1 + \alpha/r} \frac{\alpha V_{rc} \vec{V}_c}{r^2} \qquad (4.75)$$

Because of $\vec{V}_c = \vec{V}_{rc} + \vec{V}_{\varphi c}$, the formula means that when photon moves in gravitational field, it would be acted by a force in the $\vec{e}_\varphi$ direction which is vertical to the $\vec{e}_r$ direction. By using Eq.(4.52), we can write the formula above as

$$\frac{d^2 \vec{r}}{dt^2} = \left(1 - \frac{2L^2}{c^2 r^2} - \frac{\alpha L^2}{c^2 r^3}\right) \frac{2GM_0 \vec{e}_r}{(1 + \alpha/r)^3 r^2} \pm \sqrt{1 - \frac{L^2}{c^2 r^2} + \frac{\alpha L^2}{c^2 r^3}} \frac{2GM_0 L \vec{e}_\varphi}{(1 + \alpha/r)^3 c r^3} \qquad (4.76)$$

The formula can only be regarded as the motion equation of classical particle without considering the inertial mass and gravitational mass of photon which would change with speed. So it is only a formula to represent photon's acceleration, instead of dynamic equation. Though for classical particles, both are the same. But as shown below, we can obtain the proper dynamic equation of photon based on (4.75).

Let's deduce photon's motion equation under the condition $L = 0$ at first. Suppose that there is a fictitious particle corresponding to photon, of which velocity $\vec{V}'_c$ is equal to the difference between photon's velocity $\vec{c}$ in vacuum and the velocity $\vec{V}_c$ in gravitational field with $\vec{V}'_c = \vec{c} - \vec{V}_c$. When photon's velocity $V_c = c$ we have $V'_c = 0$. When photon falls down in gravitational field with speed $V_c < c$, we have $V'_c > 0$. When $V_c = 0$, we have $V'_c = c$. So similar to general particle with static mass, the fictitious particle is acted by gravitation instead of repulse force. Suppose the fictitious particle's mass is the same as the photon's equivalent static mass $m_0$, according to Eq.(4.18), when $L = 0$, the dynamic equation of fictitious particle can be written as

$$\vec{F} = \frac{d\vec{p}'}{dt} = -GM_0 m_0 \sqrt{1 - \frac{(c - V_c)^2}{c^2}} \frac{\vec{r}}{r^3} \qquad (4.77)$$

On the other hand, we define the momentum of moving photon in gravitational field as

$$\vec{p} = \frac{m_0 \vec{V}_c}{R} \qquad 4.78$$

Here $R = R(V_c)$ is an unknown function. Let $\vec{p}' \to m_0 \vec{c} - \vec{p}$ and using relation $V_c = c/(1 + \alpha/r)$ in

52

(4.77), the dynamic equation of photon would become

$$\frac{d\vec{p}}{dt} = \sqrt{\frac{2cV_c - V_c^2}{c^2}} \frac{GM_0 m_0 \vec{r}}{r^3} = \frac{\sqrt{1+2\alpha/r}}{1+\alpha/r} \frac{GM_0 m_0 \vec{r}}{r^3} \quad (4.79)$$

In general situations when $L \neq 0$, corresponding to Eq.(4.75), we have to consider the existence of force on the $\vec{e}_\varphi$ direction. So the general form of photon's dynamic equation in the gravitational field with spherical symmetry can be written as at last

$$\frac{d\vec{p}}{dt} = \frac{\sqrt{1+2\alpha/r}}{1+\alpha/r}\left(1+\frac{3L^2}{c^2 r^2}\right)\frac{GM_0 m_0 \vec{r}}{r^3} + \vec{F}_\varphi = \vec{F}_r + \vec{F}_\varphi = \vec{F} \quad (4.80)$$

Here $\vec{L} = \vec{V}_c \times \vec{r}/\sqrt{(2cV_c - V_c^2)/c^2} = \vec{V} \times \vec{r}(1+\alpha/r)/\sqrt{1+2\alpha/r}$, and the form of $\vec{F}_\varphi$ remained to be determined. In fact, it is unnecessary for as to introduce fictitious particle. We can suppose directly that the dynamic equation of photon satisfy the formula above, as long as the calculation results coincide with practical situations. Then let's determine the forms of functions $\vec{F}_\varphi$ and $R$. From (4.78) we have

$$\frac{d\vec{p}}{dt} = \frac{m_0}{R}\frac{d^2\vec{r}}{dt^2} + m_0 \vec{V}_c \frac{d}{dt}\frac{1}{R} = \vec{F} \quad (4.81)$$

So the acceleration of photon is

$$\frac{d^2\vec{r}}{dt^2} = R\left(\frac{\vec{F}}{m_0} - \vec{V}_c \frac{d}{dt}\frac{1}{R}\right) \quad (4.82)$$

Comparing (4.82) with Eq. (4.75), we get

$$R\left(\frac{\vec{F}}{m} - \vec{V}_c \frac{d}{dt}\frac{1}{R}\right) = -\frac{1}{(1+\alpha/r)^2}\frac{3\alpha L^2 \vec{r}}{2r^5} + \frac{1}{1+\alpha/r}\frac{\alpha V_{rc}\vec{V}_c}{r^2} \quad (4.83)$$

By decomposing the formula in the $\vec{e}_r$ and $\vec{e}_\varphi$ directions, we get two formulas

$$R\left(\frac{F_r}{m_0} - V_{rc}\frac{d}{dt}\frac{1}{R}\right) = -\frac{1}{(1+\alpha/r)^2}\frac{3\alpha L^2}{2r^4} + \frac{1}{1+\alpha/r}\frac{\alpha V_{rc}^2}{r^2} \quad (4.84)$$

$$R\left(\frac{F_\varphi}{m_0} - V_{\varphi c}\frac{d}{dt}\frac{1}{R}\right) = \frac{1}{1+\alpha/r}\frac{\alpha V_{rc} V_{\varphi c}}{r^2} \quad (4.85)$$

By considering $dR/dt = (dR/dr)(dr/dt) = V_{rc}(dR/dr)$, Eq.(4.84) can be rewritten as

$$\frac{dR}{dr} + P(r)R = Q(r)R^2 \quad (4.86)$$

This is the Bernoulli's equation of quasi-first order, in which

$$P(r) = \frac{1}{(1+\alpha/r)^2}\frac{3\alpha L^2}{2V_{rc}^2 r^4} - \frac{1}{1+\alpha/r}\frac{\alpha}{r^2} \quad (4.87)$$

$$Q(r) = -\frac{F_r}{m_0 V_{rc}^2} = -\frac{F_r}{m_0 c^2}\left(1+\frac{\alpha}{r}\right)^2 \quad (4.88)$$

By the integral of (4.86), we can obtain the function form of $R$, then substitute it into (4.85), we can know $F_\varphi$. So (4.85) can be written as

53

$$-R\frac{d}{dt}\frac{1}{R} = \frac{1}{1+\alpha/r}\frac{\alpha V_{rc}}{r^2} - R\frac{F_\varphi}{m_0 V_{\varphi c}} \qquad (4.89)$$

Substitute the formula into (4.84), we get

$$\vec{F}_\varphi = \left[F_r + \frac{1}{(1+\alpha/r)^2}\frac{3m_0\alpha L^2}{2Rr^4}\right]\frac{\vec{V}_{\varphi c}}{V_{rc}} = \left(F_r + \frac{3GM_0 m_0 V_{rc}^2 L^2}{Rc^4 r^4}\right)\frac{\vec{V}_{\varphi c}}{V_{rc}}$$

$$= \frac{\sqrt{1+2\alpha/r}}{1+\alpha/r}\frac{GM_0 m_0 \vec{V}_{\varphi c}}{r^2 V_{rc}}\left\{1+\frac{3L^2}{c^2 r^2}\left(1+\frac{\sqrt{1+\alpha/r}\,V_{rc}^2}{Rc^2}\right)\right\} \qquad (4.90)$$

Here $V_{\varphi c}/V_{rc} = L\sqrt{1-L^2/c^2 r^2 + \alpha L^2/c^2 r^3}/(cr)$ and $R$ is determined by Eq.(4.86).

The formulas (4.80) and (4.75) are equivalent when we calculate photon's velocity and acceleration based on them. So (4.80) can also be used to describe the three experiments to support general relativity. We call $\vec{F}_r$ as the longitudinal force and $\vec{F}_\varphi$ as the transverse force. Photons are acted by both longitudinal and transverse forces when they move in static gravitational fields with spherical symmetry. This is different from the other particles with static masses. The particles with static masses are only acted by the longitudinal force in this case.

In general situations, the integral of (4.86) is difficult. But we can do it when photon moves in the direction of radius with $L = 0$. In this case, we have $V_c = c/(1+\alpha/r)$ and get

$$F_r = \sqrt{\frac{2cV_c - V_c^2}{c^2}}\frac{GM_0 m_0}{r^2} = \frac{\sqrt{1+2\alpha/r}}{1+\alpha/r}\frac{m_0 c^2 \alpha}{2r^2} \qquad (4.91)$$

So we can define the gravitational moving mass of photon when it moves in the direction of radium

$$m_g = m_0\sqrt{\frac{2cV_c - V_c^2}{c^2}} = m_0\frac{\sqrt{1+2\alpha/r}}{1+\alpha/r} \qquad (4.92)$$

By substitute (4.91) into (4.88), then taking the integral of (4.86), we have

$$\frac{1}{R} = -\exp\left(\int P dr\right)\int Q \exp\left(-\int P dr\right) dr^{\frac{3}{2}}$$

$$= -\frac{1}{3}\left(1+\frac{\alpha}{r}\right)\left(1+\frac{2\alpha}{r}\right)^{\frac{3}{2}} = -\frac{c(2cV_c - V_c^2)^{3/2}}{3V_c^4} \qquad (4.93)$$

In this case, the momentum of photon is

$$\vec{P} = \frac{m_0 \vec{V}_c}{R} = -\frac{c(2cV_c - V_c^2)^{3/2}\vec{V}_c}{3V_c^4} \qquad (4.94)$$

The corresponding initial moving mass of photon should be define as

$$m_i = -\frac{m_0 c(2cV_c - V_c^2)^{3/2}}{3V_c^4} \qquad (4.95)$$

That is to say, the initial moving mass of photon should be a negative number. By taking the differential of (4.94), the force acted on photon is

$$F_r = \frac{dp}{dt} = \frac{m_0 c^2 \sqrt{2cV_c - V_c^2}}{V_c^3} \frac{dV_c}{dt} \rightarrow \frac{m_0 c^2 \sqrt{2cV_c - V_c^2}}{V_c^3} \frac{d^2\vec{r}}{dt^2} \tag{4.96}$$

Comparing with (4.77) and considering $V_c = c/(1+\alpha/r)$, we have

$$\frac{d^2\vec{r}}{dt^2} = \frac{1}{(1+\alpha/r)^3} \frac{2GM\vec{r}}{r^3} \tag{4.97}$$

The formula is the same as (4.76) when $L=0$, so the momentum definition of photon (4.78) is self-consistent. When $V_c \to c$ or $\alpha/r \to 0$, the acceleration of photon is double time of classical particle with static mass. the force acted on photon is repulse one, instead of gravitation. This is easy to understand. If photon is acted by gravitation, its speed would increase generally when it falls in gravitational field so that its speed would surpass light's speed in vacuum. However, this is unacceptable.

Let $U(r)$ be the potential energy of photon in the gravitational field with spherical symmetry, by considering (4.72) and (4.82), as well as $U(r) \to 0$ when $r \to \infty$, we have

$$U(r) = -\int \vec{F} \cdot d\vec{r}_r = -\int \frac{\sqrt{1+2\alpha/r}}{1+\alpha/r} \frac{m_0 c^2 \alpha}{2r^2} dr + A_3$$

$$= m_0 c^2 \left( \sqrt{1+\frac{2\alpha}{r}} - arctg\sqrt{1+\frac{2\alpha}{r}} - 1 + \frac{\pi}{4} \right) \tag{4.98}$$

In the weak field with $\alpha/r < 1$, by developing the formula into the Taylor's series, we have

$$U(r) = m_0 c^2 \left( \frac{\alpha}{2r} - \frac{3\alpha^2}{8r^2} - \frac{\alpha^3}{12r^3} - \frac{7\alpha^4}{24r^4} - \cdots \right) \tag{4.99}$$

When $\alpha/r \ll 1$, we get

$$U(r) = \frac{m_0 c^2 \alpha}{2r} = \frac{GM_0 m_0}{r} \tag{4.100}$$

In this case, photon's potential energy is the same as classical particle but takes positive value, meaning that photon is acted by repulsion force. As for photon's kinetic energy, similar to (4.29), by using (4.98), we have

$$T = \int \vec{F} \cdot d\vec{r} = \int \frac{\sqrt{1+2\alpha/r}}{1+\alpha/r} \frac{m_0 c^2 \alpha \vec{r}}{2r^3} \cdot d\vec{r}$$

$$= -m_0 c^2 \left( \sqrt{\frac{2c}{V_c} - 1} - arctg\sqrt{\frac{2c}{V_c} - 1} \right) + A_3 \tag{4.101}$$

When $V_c = c$, free photon's kinetic energy is $T_0 = m_0 c^2 = h\nu_0$, so we have $A_3 = m_0 c^2 (2 - \pi/4)$ and

$$T = m_0 c^2 \left( 2 - \frac{\pi}{4} - \sqrt{\frac{2c}{V_c} - 1} + arctg\sqrt{\frac{2c}{V_c} - 1} \right) \tag{4.102}$$

The energy conservation formula of photon in the gravitational field is

$$T + U = m_0 c^2 \left( 2 - \frac{\pi}{4} - \sqrt{\frac{2c}{V_c} - 1} + arctg \sqrt{\frac{2c}{V_c} - 1} \right)$$

$$+ m_0 c^2 \left( \sqrt{1 + \frac{2\alpha}{r}} - arctg \sqrt{1 + \frac{2\alpha}{r}} - 1 + \frac{\pi}{4} \right) = m_0 c^2 \qquad (4.103)$$

Photon can only enter the region with $T \geq 0$. For the region with $T < 0$, we have $U(r) > m_0 c^2$, or

$$\sqrt{1 + \frac{2\alpha}{r}} - arctg \sqrt{1 + \frac{2\alpha}{r}} > 2 - \frac{\pi}{4} = 1.215 \qquad (4.104)$$

In order to satisfy this formula, we should have $\alpha / r > 2.35$. In this case, photons in stars can not move along the radius direction, but they can still move around the center of gravitation with $L \neq 0$. So the lights emit by star with $r < \alpha / 2.35$ can not be seen by the observers outsider the star. This kind of star can be considered as black hole. But this black hole has no singularity. The event horizon $r = \alpha / 2.35$ can also be regarded as potential barrier with height $U(r) = m_0 c^2$ which photons can not pass through. When photon reaches the potential barrier along the radius direction with speed $V_c = c / (1 + 2.35) \approx 0.3c$, it would be rebounded. The problem of gravitational redshift of light will be discussed in Section 7 again.

As mentioned before that the Einstein's theory of gravity can not be a universal one. The effectiveness of the Schwarzschild solution can only be considered as a haphazard but excellent coincidence in weak field. We should reestablish gravitational theory in flat space-time based on the spherical symmetry solution the Einstein's equation. In fact, we can consider (4.25) and (4.80) as the basic dynamic equations of particles moving in the gravitational fields caused by other static particles. Then by means of the method of force's superposition, we can construct gravitational interactions among the bodies with any different forms. No any other gravitational equations in curved space-time are needed again. In this way, we can establish the general theory of gravity in the form of electromagnetic interaction.

## 6. Gravitational redshift and the revised formula of Doppler shift in gravitational field

According to general relativity, gravitational field would cause time delay so that light's frequency becomes small and light's wave length become longer. But as shown in Second 3, this kind of explanation has a logical difficulty. The current explanation of gravitational redshift is based on the hypothesis that light's speed is unchanged in gravitational field. It is just this hypothesis to cause the difficulty. Strictly based on the Schwarzschild solution of the Einstein's equation of gravitational field, for the observers who are at rest in the gravitational field, light's speed is less than its speed in vacuum when it moves along the radius direction of gravitational with spherical symmetry. If considering the fact that light's speed, wave length and frequency change simultaneously, we would not have any difficulty again to explain the gravitational redshift.

Let the proper wave length and frequency of a certain atomic be $\lambda_1$ and $\nu_1$ without gravitational field in space, we have $c = \lambda_1 \nu_1$ and $h\nu_1 = m_1 c^2$. Then we place the atom into the point $r$ in the gravitational field with spherical symmetry. Owing to the gravitational interaction, the energy levels of atom would change. Suppose the vibration frequency of atom or the frequency of light emitted by the atom becomes $\nu$, but light's wave length is still unchanged in this case. That is to say, light's wave length is still $\lambda_1$, so light's speed becomes $V = \lambda_1 \nu$. Because photon's potential energy is $U(r)$ in gravitational field, so photon's energy becomes $h\nu + U(r)$. When the photon reaches to the observers outsider the gravitational field with $r \to \infty$, its wave length and frequency becomes $\lambda_0$ and $\nu_0$. We have $c = \lambda_0 \nu_0$ and the law of energy conservation

$$h\nu + U(r) = h\nu_0 = m_0 c^2 \qquad (4.105)$$

It is obvious that $h\nu$ corresponds photon's dynamic energy in gravitational field. That is to say, we can suppose that photon's frequency is only relative to its frequency, having nothing to do with its potential energy. The result indicates that if the energy of light emitted by an atom outsider the gravitational field is $E = m_1 c^2$, when the atom is putted into the gravitational field, the energy of light emitted by the same atom would become $E_0 = m_0 c^2$ when the light arrives outsider the field. We can have $E_1 > m_0 c^2 = E_0$. The reason would be that the energy level of atom becomes small so that littler energy is need to staminate atom to emitting photon. In this way, owing to the relation (2.19) when atomic clocks are used to measure time, the clocks located in gravitational field would become slow comparing with the locks located outsider the field. The result coincides with the experiments of atomic clocks cincturing the earth. But as mentioned in Second 2, if mechanical clocks are used in the experiments, we can not find the time delay caused by gravitation. This is different from that in general relativity.

On the other hand, light's frequency is an unmeasured quantity. In the practical experiments of red shift of spectrum, what measured directly is light's wave length. We should use wave length to define the redshift gravitation. Suppose that the wave length of light emitted by an atom located in the point $r_1$ of gravitational field is $\lambda_1$ with $V_c = \lambda_1 \nu$. The wave length accepted by observers outsider the field is $\lambda_0$. For the consistence with the definition of the Doppler shift, by considering (4.105), we also define gravitational redshift as

$$Z_g = \frac{\lambda_0}{\lambda_1} - 1 = \frac{c}{V_c}\frac{\nu}{\nu_0} - 1 = \left(1 + \frac{\alpha}{r_1}\right)\frac{h\nu_0 - U(r_1)}{h\nu_0} - 1 \qquad (4.106)$$

For the gravitational field with spherical symmetry, substituting (4.98) into (4.106), we get

$$Z_g = \left(1 + \frac{\alpha}{r_1}\right)\left(2 - \frac{\pi}{4} - \sqrt{1 + \frac{2\alpha}{r_1}} + arctg\sqrt{1 + \frac{2\alpha}{r_1}}\right) - 1 \qquad (4.107)$$

When $\alpha/r_1 = 1$, we have $Z_g = 0.5$. When $\alpha/r_1 \leq 1$, we can developing the formula into Taylor's series

$$Z_g = \left(1 + \frac{\alpha}{r_1}\right)\left(1 - \frac{\alpha}{2r_1} + \frac{3\alpha^2}{8r_1^2} + \frac{\alpha^3}{12r_1^3} + \frac{7\alpha^4}{24r_1^4} + \cdots\right) - 1 \qquad (4.108)$$

It is known that when $\alpha/r_1 < 1$, we have $Z > 0$. In weak field with $\alpha/r_1 \ll 1$, we have

$$Z_g = \left(1 + \frac{\alpha}{r_1}\right)\left(1 - \frac{\alpha}{2r_1}\right) - 1 = \frac{\alpha}{2r_1} = \frac{GM_0}{c^2 r_1} \qquad (4.109)$$

The result coincides with experiments.

For the same reason, when light's source moves in gravitational field, the Doppler formula of frequency shift should be revised. Let $K_1$ represent the reference frame in which the light's source is at rest and light's (proper) wave length and (proper) frequency are $\lambda_1$ and $\nu_1$. When light's source moves in a uniform speed $V$ relative to the reference frame $K_0$, the wave length and frequency observed by the observer become $\lambda_0$ and $\nu_0$. According to the invariable principle of phase in special relativity, we obtain the Doppler relation between two frequencies

$$\nu_1 = \frac{\nu_0(1 - V\cos\theta_0/c)}{\sqrt{1 - V^2/c^2}} \tag{4.110}$$

When light's moves away from the observers, we take $\theta_0 = \pi$ and get the redshift

$$Z_d = \frac{\lambda_0}{\lambda_1} - 1 = \frac{\nu_1}{\nu_0} - 1 = \sqrt{\frac{1 + V/c}{1 - V/c}} - 1 \tag{4.111}$$

However, this formula is based on the invariability principle of light's speed, so it is only suitable for the light's source moving in free space. When the light's source moves in gravitational field, the formula should be revised.

Suppose that the center mass of gravitational field with spherical symmetry is $M_0$ and the reference frame $K_1$ moves in speed $V$ along the direction of radius direction of gravitational field, relative to the resting reference frame $K_0$. The wave length and frequency of light emitted by the light's source which at rest in $K_1$ is $\lambda_1$ and $\nu_1$. The wave length and frequency measured by the observers who are at rest in the reference frame $K_0$ outsider the gravitational field are $\lambda_0$ and $\nu_0$. For the observer who are at rest at $K_1$, in light of (4.70), the motion equation of photon in gravitational field is

$$\frac{dr_1}{dt_1} = \frac{c}{1 + \alpha/r_1} \qquad \text{or} \qquad ct_1 = r_1 + \alpha \ln r_1 \tag{4.112}$$

Let $l_1 = r_1 + \alpha \ln r_1 = r_1(1 + \alpha \ln r_1/r_1) = r_1 f(r_1)$, we have $ds^2 = dl_1^2 - c^2 dt_1^2 = 0$. Because $ds = 0$ is an invariable quantity for light's motion, the result means that when light moves along radius direction in gravitational field, if we use $l_1$ to substitute $r_1$, photon can be considered to move in free space without gravitation. According to the invariable principle of phase $\Phi$ in special relativity, we have

$$\Phi_0 = \frac{\omega_0}{c}(ct_0 - l_0 \cos\theta_0) = \frac{\omega_1}{c}(ct_1 - l_1 \cos\theta_1) = \Phi_1 \tag{4.113}$$

In this way, we can get the relation of transformation

$$\omega_1 = \frac{\omega_0(1 - V\cos\theta_0/c)}{\sqrt{1 - V^2/c^2}} \qquad \cos\theta_1 = \frac{\cos\theta_0 - V/c}{1 - V\cos\theta_0/c} \tag{4.114}$$

By relation $\omega = 2\pi\nu$, we can also get (4.110) from (4.114). Meanwhile, because we have define $l_0 = r_0 f(r_0)$ and $l_1 = r_1 f(r_1)$, we also have the Lorentz transformation

$$r_1 f(r_1) = \frac{r_0 f(r_0) - Vt_0}{\sqrt{1 - V^2/c^2}} \qquad t_1 = \frac{t_1 - Vr_1 f(r_1)/c}{\sqrt{1 - V^2/c^2}} \tag{4.115}$$

From the first formula, we can obtain $r_1 = F(r_0, t_0)$. So it is obvious that when ligh't source moves along the radius direction in gravitational field, the Doppler formula of frequency transformation is unchanged, what is changed is the Lorentz transformation relations between space-time coordinates $r_1, r_0, t_1$ and $t_0$.

Similarly, because light's frequency can not be measured directly, we should use wave length to represent redshift. Suppose that light is emitted from the surface of a celestial body with radius $r_1$, light's speed becomes $V_c = \lambda_1 \nu_1 = c/(1 + \alpha_1/r_1)$ in the gravitational field. The light's wave length and frequency observed by the observer outsider the gravitational field become $\lambda_0$ and $\nu_0$ with $c = \lambda_0 \nu_0$. So by considering gravitational influence on light's speed, the Doppler formula become

$$Z_d = \frac{\lambda_0 - \lambda_1}{\lambda_1} = \frac{c}{V_c}\frac{v_1}{v_0} - 1 = \left(1 + \frac{\alpha}{r_1}\right)\sqrt{\frac{c+V}{c-V}} - 1 \qquad (4.116)$$

In which $\alpha/r_1$ is the revised factor caused by gravitation.

## 7. The test to verify the change of light's speed caused by the earth gravitational field

Because the conclusion that light's speed would change in the gravitational field is a very important, we need direct verifications for this result. Based on formula (4.52), we can propose a test to verify this prediction in the gravitational field of the earth. The test can be considered as a new verification of general relativity in weak gravitational field. Because under the condition $\alpha/r \ll 1$, (4.52) is consistent with (4.69) in this paper, the conclusion is suitable for this paper's theory, though in strong fields, both are different. So we discuss this problem only based on formula (4.52) below.

In light of special relativity, light's speed is a constant with isotropy in vacuum without the existence of gravitation or interaction. But it is still an uncertain problem in the current theory whether or not light's speed would change in gravitational field. Because there is interaction between photon and material in gravitational field, photon would not be free one. It is a rational speculation that light's speed would change in gravitational field. As shown below, we prove this point based on the Einstein's theory of gravitation. On the other hand, according to the common understanding of general relativity, we have both definitions of the coordinate time and the proper time, as well as the coordinate length and the proper length. Proper time and proper length are also called as stander clock and stander ruler. Because coordinate time and coordinate length are variable in the different space-time points in gravitational field, it is meaningless for us to use them do any measurement. When we measure time, space and object's motions in gravitational field, what can be used are only stander clock and stander ruler according to the current theory. It has been proved that in a static gravitational field, when stander clock and stander ruler are used, light's speed, equal to its speed in vacuum, is still a constant[1]. Therefore at present, in the concrete calculations of astrophysical and cosmological problems, as well as in the theoretical explanations of astronomical observations, we always take light's speed as a constant.

On the other hand, as we known that only on the locally inertial reference frame, we can define stander ruler (proper length) and stander clock (proper time). The locally inertial reference frame is considered to be one falling freely in gravitational field, in which the Schwarzschild metric becomes that of flat space-time with the form of $ds^2 = c^2 d\tau^2 - dR^2$. So only on the locally inertial reference frame with stander ruler and stander clock, light's speed would be a constant equal to its speed in vacuum. But for observers at rest in the gravitational field of the earth which is not a locally inertial reference frame, light's speed could not be a constant, because no stander ruler and stander clock can be defined according to the current understanding.

In fact, in general relativity, when we calculate the perihelion precession of the Mercury, the deviation of light and the delay of radar wave in the solar gravitational field, we actually use coordinate time and coordinate length, in stead of proper time and proper length. The calculating results of the deviation of light and the delay of radar wave in the gravitational field of the sun are

$$\theta = \frac{4GM_0}{c^2 r_s} \qquad\qquad \Delta t = \frac{4GM_0}{c^3}\left(1 + \ln\frac{4rr'}{r_s}\right) \qquad (4.117)$$

Here $r_s$ is the solar radius, $r$ and $r'$ are the distances between the earth, the Mercury and the sun. It can be seen from the deduction processes of (4.46) that the space-time coordinates $r, r', r_s, \theta$ and $t$ in the formula are the same as defined in formula (4.117). By comparing the results of (4.117) with experiments directly, we claim that general relativity is supported by experiments. However, we should be clear in mind that the formula (4.117) is based on the static reference frame of the sun, in which coordinate time and coordinate length are used. We have not transformed them into proper time and proper length. Meanwhile, it should be noticed that all experiments and observations are carried out on the earth reference frame, in stead of that on the sun. Only because relative velocity between them is small, the effect of special relativity is neglected. The availability of (4.117) comparing with practical experiments indicates that and the formula represents actually the results described in flat space-time. It is an accidental consistence for the Einstein's theory of gravitation described in curve space-time. Only the solution of the Schwarzschild solution is effective in weak field. From this point, it can be seen why we should establish the theory of gravity in flat space-time.

Therefore, for the same problem about light's motion in gravitational field, we can also verify the effect of the earth's gravitational field on the light's speed and its isotropy by experiments. The method of the Michelson—Morley interference is used for this purpose. When light moves along the vertical direction of the earth's surface, we have $L = 0$. In this case, according to (4.52), light's vertical speed is

$$V_{c\perp} = c\left(1 - \frac{\alpha}{r}\right) \tag{4.118}$$

Here $c$ is light's speed when $r \to \infty$. When light moves along the parallel direction of the earth's surface, we have $L = cr_e$. Here $r_e$ is the earth's radius. We have $\alpha / r_e = 1.39 \times 10^{-9} \ll 1$ for the earth. So according to (4.52), light's parallel speed on the earth's surface is

$$V_{c11} = c\left(1 - \frac{\alpha}{r_e}\right)\sqrt{1 + \frac{\alpha}{r_e}} \approx c\left(1 - \frac{\alpha}{2r_e}\right) \tag{4.119}$$

If let $V_r = 0$ in (10), we get $L/(cr) = (1 - \alpha/r)^{-1/2}$. Then substitute the result into (11), we can also get the same result. Because we always measure light's speed on the earth's surface, according to the current value, we can take $c_{11} = 2.997924580 \times 10^8 m/s$. Therefore, on the earth's surface, we have

$$V_{c\perp} = \frac{V_{c11}}{\sqrt{1 + \alpha/r_e}} = c_{11}(1 - 6.95 \times 10^{-10}) = 2.997924578 \times 10^8 m/s \tag{4.120}$$

We have $V_{c11} - V_{c\perp} = 0.2$. The light's vertical speed is less than its parallel speed on the earth's surface. Thought the change of light's speed and the violation of isotropy are very small, we can still verify their existence by using the method of the Michelson—Morley interference. Suppose that the Michelson interferometer's two arms have the same length $h$. One arm is vertical to the earth's surface and another is parallel to the surface. In this way, there exists the difference of gravitational potential between them. When a bundle of light moves along the vertical direction of the earth's surface from the coordinate point $r_e$ to the point $r_e + h$ with $h \ll r_e$, spent time is

$$\Delta t_1 = \int_{r_e}^{r_e+h} \frac{dr}{V_{c\perp}} = \int_{r_e}^{r_e+h} \frac{dr}{c(1-\alpha/r)} = \frac{1}{c}\left[h + \alpha \ln\left(1 + \frac{h}{r_e - \alpha}\right)\right] \approx \frac{h}{c}\left(1 + \frac{\alpha}{r_e}\right) \qquad (4.121)$$

While a bundle of light moves $h$ distance along the parallel direction of the earth's surface, spent time is

$$\Delta t_2 = \frac{h}{c_{11}} \approx \frac{h}{c}\left(1 + \frac{\alpha}{2r_e}\right) \qquad (4.122)$$

So when two lights are reflected back, the time difference is

$$\Delta T = 2(\Delta t_1 - \Delta t_2) = \frac{\alpha h}{c r_e} \qquad (4.123)$$

When the interferometer's arms are turned over $90^0$, the time deference becomes

$$\Delta T' = -2(\Delta t_1 - \Delta t_2) = -\frac{\alpha h}{c r_e} \qquad (4.124)$$

So before and after the interferometer's arms are tuned, the change of time difference is

$$\delta t = \Delta T - \Delta T' = \frac{2\alpha h}{\lambda r_e} \qquad (4.125)$$

We take light's wave length $\lambda = 4 \times 10^{-7} m$ and $h = 10m$ in the experiment. So after interferometer's arms are tuned over $90^0$, the shift of interference stripe would be

$$\Delta = \frac{c}{\lambda}\delta t = \frac{2\alpha h}{\lambda r_e} = 6.95 \times 10^{-2} \qquad (4.126)$$

Because the shift of $10^{-2}$ stripe can be observed for the precise Michelson interferometer[2], this is a directly observable result. If the shift of interference stripe can be found, it would mean that light's speed changes and its isotropy is violated in the gravitational field of the earth. At present, all experiments of the Michelson—Morley interference used to prove light's speed invariable are carried out on the earth's surface. In this situation, the interferometer's two arms are parallel to the earth's surface without the difference gravitational potential. This may be one of the reasons that the shift of interference stripe can not be found in the experiments. If the difference of gravitational potential exists in the experiments, it may be possible for us to observer the shift of interference stripe.

  Because the formula (4.52) is deduced from the Schwarzschild solution of the Einstein's equation of gravitational field, the experiment can be regarded as a new verification for general relativity in the weak gravitational field. It should be emphasized that all three results shown in formulas (4.104) and (4.52) are deduced from the formula (4.46). Now that the effects shown in (4.117) have been verified to exist, the effect shown in (4.52) would also exist. Thought the change of light's speed in the earth's gravitational field is very small, it would be great in strong field. For example, we have $\alpha/r = 2.12 \times 10^{-6}$ on the surface of the sun. So for the sun, we would have $V_{c11} - V_{c\perp} \approx 300 m/s$. This is big value though the gravitational field of the sun is still not very strong. Under limit condition, for the so-called black hole with $\alpha/r \to 1$, we would have light's speed $V \to 0$ in light of (4.52). But according to the current understanding, in this case, we still have $\lambda \nu = c \neq 0$ with light's wave length $\lambda \to \infty$ and frequency $\nu \to 0$. The results are completely different. At present, light's speed is always regarded as a constant in the concrete calculations of astrophysical and cosmological problems, as well as in the theoretical explanations of astronomical observations. If the experiment shows that light's speed would change in gravitational field, the result would produce great effects on foundational physics, astrophysics and cosmology. Conversely, if the shift of interference stripe can not be observed in the experiment, we should ask why the prediction of general relativity can not coincide with practice. So no matter whether or not the shift of interference stripe can be caused, the experiment is meaningful and worthy to be done seriously.

## Section 5   Gravitational Theory Established in Flat Space-time

**1. Introduction**

As we shown before that it is improper to consider the Einstein's theory of gravity as the foundational interaction theory of gravity. The real value of the Einstein's theory is to provide the Schwarzschild metric of static gravitational field with spherical symmetry. It is useful to describe object's motions in the weak field, though it may be an accidental coincident. The later theory should be consistent with it. After the metric is transformed into flat space-time for discussion, the dynamic equation of gravitational interaction between two particles can be obtained. By considering the similarity between classical electromagnetic and gravitational theories as well as introducing magnetic-like gravitation, we can establish a more rational gravitational theory with the Lorentz invariability. The descriptions of electromagnetic and gravitational interactions can also become consistent.

**2. The gravitational theory between two objects with static masses**

In Second 1  we prove that the absolutely resting reference frame should exist. In the absolutely resting reference frame, an object's mass would be smallest. So we should establish gravitational theory in the absolutely resting reference frame firstly. After that, we transform the theory into other inertial reference frames for discussion. Let $m_{i0}$ represent the static mass of a particle in the absolutely resting reference frame, $\vec{V}$ represent its velocity relative to the absolutely resting reference frame, $m_i$ represents the inertial mass of particle. According to special relativity, we have $m_i = m_{i0}/\sqrt{1-V^2/c^2}$. As shown before, when a particle moves in a gravitational field caused by a static object with spherical symmetry and static mass $M$, the gravitation acted on the particle is

$$\frac{d\vec{p}}{dt} = -GMm_0\left(1 + \frac{3L^2}{c^2r^2}\right)\sqrt{1-\frac{V^2}{c^2}}\frac{\vec{r}}{r^3} \qquad (5.1)$$

Here $\vec{p} = m_i d\vec{r}/dt$, $L$ is a constant. Based on the formula, we can definite gravitational moving mass

$$m_g = m_{g0}\sqrt{1 - \frac{V^2}{c^2}} \qquad (5.2)$$

Here $m_{g0}$ is gravitational static mass. Because the *Eötvös* type of experiments has shown that gravitational static mass is equivalent with inertial static mass, we have $m_{i0} = m_{g0} = m_0$. But in general situations when $\vec{V} \neq 0$, gravitational moving mass is not equivalent with inertial moving mass with $m_i \neq m_g$. Gravitational moving mass is biggest when object's speed is zero. When object's speed reaches light's speed, its gravitational moving mass becomes zero. The situation is just opposite to inertial mass. The result is interested that general relativity is based on the equivalent principle between gravitational mass and inertial mass. But it only indicates that gravitational static mass is equivalent with inertial static mass actually. After the Schwarzschild solution of the Einstein's theory of gravity is transformed into flat space-time for description, we reach the result that gravitational moving mass is not equivalent with inertial moving mass.

The classical Newtonian theory of gravitation describes the gravitation between two static objects actually. The theory has two defects. One is that it can not satisfy the Lorentz invariability. Another is that it can not describe small effects of gravitation such as the perihelion precession of the Mercury and so on. So we have to revise it from these two sides. By considering the fact that the validity of electromagnetic interaction theory and the comparability between the Newtonian formula of gravitation and the Coulomb formula of static electrics, if there exists unity between gravitational and electromagnetic interactions, gravitation would take the similar form of electromagnetic force, instead of that electromagnetic interaction should be coincide with gravitation described in curved space-time. It is well known that too many singularities appear in the Einstein's theory of gravitation described in curved space-time.

Therefore, we introduce the concepts of electric-like and magnetic-like gravitations. Suppose that there are two particles with static masses $m_{10}$ and $m_{20}$ moving in velocities $\vec{V}_1$ and $\vec{V}_2$ individually relative to the absolutely static reference frame. The electric-like gravitation, which is caused by the particle with static mass $m_{20}$ and acted on the particle with static mass $m_{10}$ and, is defined as

$$\vec{F}_e = -\frac{Gm_{g1}m_{g2}\vec{r}}{r^3}\left(1 + \frac{3L_1^2}{c^2 r^2}\right) = \frac{m_{g1}m_{g2}\vec{r}}{4\pi\varepsilon_g r^3}\left(1 + \frac{3(\vec{V}_1 \times \vec{e}_r)^2}{c^2\sqrt{1 - V_1^2/c^2}}\right) \qquad (5.3)$$

Here $m_{g1}$ and $m_{g2}$ is defined in Eq.(5.3), $\vec{r} = \vec{r}_1 - \vec{r}_2$, $\vec{e}_r = \vec{r}/r$, $\varepsilon_g$ is the so-called gravitational electric-like dielectric constant

$$\varepsilon_g = -\frac{1}{4\pi G} \qquad (5.4)$$

If there are only two particles in the system, the angle momentum of unit mass $L_1 = |\vec{V}_1 \times \vec{r}|/\sqrt{1 - V_1^2/c^2}$ is a constant. If there are more particles in the system, $\vec{L}_1$ is not a constant in general. Meanwhile, similar to electromagnetic theory, we define the magnetic-like gravitation as

$$\vec{F}_m = \frac{\mu_g m_{g1} m_{g2}}{4\pi}\frac{\vec{V}_1 \times (\vec{V}_2 \times \vec{r})}{r^3} = \frac{\mu_g}{4\pi}\frac{\vec{J}_{g1} \times (\vec{J}_{g2} \times \vec{r})}{r^3} = \vec{J}_{g1} \times \vec{B}_{g2} \qquad (5.5)$$

$$\vec{J}_{gi} = m_{gi}\vec{V}_i = m_{i0}\vec{V}_i\sqrt{1 - \frac{V_i^2}{c^2}} \qquad (5.6)$$

Here $\vec{J}_{gi}$ is the density of mass flow, $\mu_g$ is the gravitational magnetic-like permeability and $\vec{B}_g$ is the intensity of magnetic-like gravitational field

$$\vec{B}_g = \frac{\mu_g}{4\pi} \frac{\vec{J}_g \times \vec{r}}{r^3} \tag{5.7}$$

Similar to electromagnetic theory, we can deduce

$$\frac{1}{\sqrt{\varepsilon_g \mu_g}} = c \tag{5.8}$$

By means of Eq.(5.4), we get

$$\mu_g = -\frac{4\pi G}{c^2} \tag{5.9}$$

It is useful to compare the intensities of magnetic-like and electromagnetic gravitations. We have

$$F_m \sim \frac{\mu_g}{4\pi} \frac{J_{g1} J_{g2}}{r^2} \sim \frac{G}{c^2} \frac{m_{g1} m_{g2} V_1 V_2}{r^2} \sim F_e \frac{V^2}{c^2} \tag{5.10}$$

So when $V \ll c$, the magnetic-like gravitation can be neglected comparing with the electric-like gravitation. This is just the reason why the Newtonian theory of gravitation is quite effect without considering magnetic-like gravitations. In electromagnetic interaction, charged particle's speeds are great in general so that strong magnetic phenomena would be caused. But in the strong gravitational field with particle's speed $V \sim c$, magnetic-like gravitations can not be neglected.

In this way, we can establish the Maxwell's equations of gravitational fields in the similar form of electromagnetic theory. For a particle with static mass $m_0$ and velocity $\vec{V}$, we define the intensity of its electric-like gravitational field as

$$\vec{E}_g = \frac{m_g \vec{r}}{4\pi \varepsilon_g r^3} = \frac{m_0 \sqrt{1-V^2/c^2}}{4\pi \varepsilon_g} \frac{\vec{r}}{r^3} \tag{5.11}$$

$\vec{E}_g$ is relative to particle's speed. This is different from the intensity of electric field. When material's mass is distributed continuously, the density function of gravitational moving mass should de defined as

$$\rho_g(\vec{r},t) = \rho_0(\vec{r},t) \sqrt{1 - \frac{V(\vec{r},t)^2}{c^2}} \tag{5.12}$$

In which $\rho_0(\vec{r},t)$ is the density distributive function when material is at rest. In this case, Eq.(5.11) should be rewritten as

$$\vec{E}_g(\vec{x},t) = \frac{1}{4\pi \varepsilon_g} \int \frac{\rho_g(\vec{x}',t) \vec{r}}{r^3} d^3 \vec{x}' \tag{5.13}$$

Here $\vec{r} = \vec{x} - \vec{x}'$. Similarly, we have

$$\nabla \cdot \vec{E}_g(\vec{x},t) = \frac{\rho_g}{\varepsilon_g} \tag{5.14}$$

For the intensity of magnetic-like gravitational field, in this case, we also have

$$\vec{B}_g(\vec{x},t) = \frac{\mu_g}{4\pi} \int \frac{\vec{J}_g(\vec{x}',t) \times \vec{r}}{r^3} d^3 \vec{x}' \tag{5.15}$$

In which

$$\vec{J}_g(\vec{x}',t) = \rho_g(\vec{x}',t)\vec{V}(\vec{x}',t) = \rho_0(\vec{x}',t)\sqrt{1-\frac{V^2(\vec{x}',t)}{c^2}}\vec{V}(\vec{x}',t) \qquad (5.16)$$

Also we have

$$\nabla \cdot \vec{B}_g(\vec{x},t) = 0 \qquad (5.17)$$

Similar to electromagnetic theory, suppose that there exist the law of induction between electric-like and magnetic-like gravitational fields

$$\nabla \times \vec{E}_g = -\frac{\partial \vec{B}_g}{\partial t} \qquad \nabla \times \vec{B}_g = \mu_g \vec{J}_g + \mu_g \varepsilon_g \frac{\partial \vec{E}_g}{\partial t} \qquad (5.18)$$

The formulas (5.14), (5.17) and (5.18) are the Maxwell's equation set of gravitational fields. In the formulas, the forms of $\vec{E}_g$ and $\vec{B}_g$ are determined by $\rho_g(\vec{x},t)$ and $\vec{J}_g(\vec{x},t)$. Comparing with electromagnetic theory, the only difference is that there is contraction factor of relativity in the mass density $\rho_g(\vec{x},t)$ and the mass flow density $\vec{J}_g(\vec{x},t)$. Therefore, this kind of gravitational theory is obviously invariable under the Lorentz transformation. The descriptions of gravitational and electromagnetic interactions also become consistent.

On the other hand, by means of the intensions of electric-like and magnetic-like gravitational fields, when a particle with gravitational moving mass $m'_g$ and velocity $\vec{V}'$ moves in the gravitational field caused by another particle with gravitational moving mass $m_g$ and velocity $\vec{V}$, the Lorentz force acted on the first particle can be represented as

$$\vec{F} = m'_g\left[\left(1+\frac{(\vec{V}'\times\vec{e}_r)^2}{c\sqrt{c^2-V'^2}}\right)\vec{E}_g + \vec{V}'\times\vec{B}_g\right] \qquad (5.19)$$

Comparing with the Lorentz formula of electromagnetic theory, there exist an additional item relative to angle momentum, besides the differences of gravitational moving mass density and mass flow density. If we further define longitudinal gravitational moving mass

$$m'_{gL}(V') = m'_g\left(1+\frac{(\vec{V}'\times\vec{e}_r)^2}{c\sqrt{c^2-V'^2}}\right) \qquad (5.20)$$

the Lorentz formula of gravitation can be written as more simple form

$$\vec{F} = m'_{gL}\vec{E}_g + m'_g\vec{V}'\times\vec{B}_g \qquad (5.21)$$

The form is completely the same as that of electromagnetic interaction, except that the definitions of gravitational moving masses $m'_{gL}$ and $m'_g$ are different from the inertial moving masses in electromagnetic theory.

## 2. The gravitational theory between objects with static masses and photons

The expression of photon's gravity is discussed below. In the spherical coordinate system, the unit vectors of directions are $\vec{e}_r$, $\vec{e}_\theta$ and $\vec{e}_\varphi$ with $\vec{e}_\varphi = \vec{e}_r \times \vec{e}_\theta = \vec{r}\times\vec{e}_\theta/r$. So Eq.(4.87) can be written as

$$\vec{F}_\varphi = \sqrt{\frac{2cV-V^2}{c^2}}\frac{GMm_0V_\varphi(\vec{r}\times\vec{e}_\theta)}{r^3 V_r}\left\{1+\frac{3(\vec{V}\times\vec{e}_r)^2}{c\sqrt{2cV-V^2}}\left(1+\frac{V_r^2}{Rc\sqrt{2cV-V^2}}\right)\right\} \qquad (5.22)$$

For photons, we can also introduce the longitudinal gravitational moving mass $m_{gL}$ and transverse gravitational moving mass $m_{gT}$ with

$$m_{gL} = \frac{m_0 \sqrt{2cV_c - V_c^2}}{c}\left(1 + \frac{(\vec{V}_c \times \vec{e}_r)^2}{c\sqrt{2cV_c - V_c^2}}\right) \quad (5.23)$$

$$m_{gT} = \frac{m_0 V_{\varphi c}\sqrt{2cV_c - V_c^2}}{cV_{rc}}\left\{1 + \frac{3(\vec{V}_c \times \vec{e}_r)^2}{c\sqrt{2cV_c - V_c^2}}\left(1 + \frac{\vec{V}_{rc}^2}{R(V_c)c\sqrt{2cV_c - V_c^2}}\right)\right\} \quad (5.24)$$

So when a photon moves in a spherically symmetrical gravitational field, the electric-like gravitational force acted on the photon can be written as

$$\vec{F}_e = \vec{F}_r + \vec{F}_\varphi = m_{gL}\vec{E}_g + m_{gT}\left(\vec{E}_g \times \vec{e}_\theta\right) \quad (5.25)$$

In which $\vec{E}_g$ is determined by Eq.(5.11). If the center mass has a moving velocity, the corresponding magnetic-like gravitation should be added. Because the sun's velocity is very small, according to Eq.(5.10) with $F_m / F_e \sim V/c$, the magnetic-like gravitation can be neglected. So in the weak field of the sun, photon's acceleration can still be represented by Eq.(4.74). In general situations, when a particle moves in the gravitational field, the Lorentz formula of gravitation can be uniformly written as

$$\vec{F}_g = m_{gL}\vec{E}_g + m_{gT}\left(\vec{E}_g \times \vec{e}_\theta\right) + m_g \vec{V} \times \vec{B}_g \quad (5.26)$$

What we discuss above is the theory described in the absolutely resting reference frames. By the Lorentz coordinate transformation and velocity transformation in special relativity, we can describe the theory in another inertial reference frame moving relative to the absolutely resting reference frame. We discuss the transformation of gravitational moving mass below. Suppose that the sun's absolute velocity is $\vec{V}_2$ and the planet's absolute velocity is $\vec{V}_1$ relative to absolutely resting reference frame individually. Because these two velocities are small, for simplification, we only use the Galilei's ruler of velocity transformation. So the planet's velocity relative to the sun can be written as $\vec{V} = \vec{V}_1 - \vec{V}_2$. When the sun is considered at rest, its static mass can be considered to be equivalent to $M'_0 = M_0\sqrt{1 - V_2^2/c^2} = M_g$ and the planet's static mass can also be considered to be equivalent to $m'_0$. So the planet's gravitational moving mass can be considered to be equivalent to

$$m_g = m'_0\sqrt{1 - \frac{V^2}{c^2}} \qquad m'_0 = m_0 \frac{\sqrt{1 - V_1^2/c^2}}{\sqrt{1 - V^2/c^2}} \quad (5.27)$$

Then let $m_{g2} = M_g$, $m_{g1} = m_g$ and $\vec{V}_1 = \vec{V} + \vec{V}_2$ in (5.2), we can use it to represent the approximate formula of electric-like gravitation in the reference frame in which the sun is considered at rest. In this case, the sun's magnetic-like gravitational can be regarded as zero approximately. Of cause, if two object's speeds are great, the addition formula of velocity in special relativity and the magnetic-like gravitation should be taken into account.

Meanwhile, this kind of gravitational theory has some natures below.

1. In this theory, quantization of gravitational field can be carried out in the similar form of electromagnetic field. Photon's spin is 1 instead of 2.

2. Similar to electromagnetic theory, this gravitational theory may be renormalizable. So it may provide a simplest foundation for the unified theory of four forces.

3. The energy momentum tensors of gravitational fields can be also defined well as that done in the electromagnetic fields. The difficulty existing in general relativity can be avoided.

4. There exist dipole radiations of gravitational waves in this theory as that in electromagnetic theory.

According to general relativity, the lowest order of gravitational radiation is the fourth order. There exist no dipole radiations. This point is one of biggest differences between two theories, which can be used to decide which one is alight. At present, we only use quadrupole resonance apparatus to detect gravitational waves but find nothing. It may be more effective to use dipole resonance apparatus to do it.

It is useful to estimate the radiation strength of gravitational wave in the theory. Similar to electromagnetic theory, when a particle with static mass $m$ moves in a speed $V << c$, the power of its gravitational radiation is $P_g = m^2 a^2 /(6\pi\varepsilon_g c^3)$. Here $a$ is particle's acceleration. So for an electron, the ratio of electromagnetic radiation and gravitational radiation is the same as that in general relativity with $P_e / p_g \approx e^2 \varepsilon_g /(m^2 \varepsilon_e) \approx 4 \times 10^{44}$.

# Section 6  Cosmological Theory Established in Flat Space-time

## 1. Introduction

It is proved that according to the Robertson---Walker metric, the velocity of light which is emitted by the celestial bodies in the expansive universe would obey the Galileo's addition ruler in classical mechanics, in stead of the Einstein's addition ruler in special relativity. This result violates the principle of invariance of light's velocity and contradicts with physical experiments and astronomical observations. So the Robertson---Walker metric can not be used as the basic frame of space-time to describe the expensive universe. The theoretical foundation of modern cosmology must be reestablished. Some established viewpoints and conclusions should be re-surveyed. A most direct result is that the affirmation about dark energy governing our universe and the universal accelerating expansion deduced from the fit between cosmological theory and the observations of high red-shift Ia type supernovae are untrue. When the revised Newtonian formula of gravitation is used to discuss the problem of the universal expansion, the new motion equation of cosmology is obtained which is similar to the Friedmann equation of cosmology without containing cosmic constant. By the formula, the Hubble diagrams of high red-shift Ia type supernovae can be explained well without using the concept of dark energy. The problem of cosmic constant which has puzzled physical circle for a long tome can be get rid of thoroughly. Meanwhile, we do not need to suppose more than three fourth of total material in our universe to be non-baryon dark material, if they exist indeed. The problem of the universal age can also be solved well.

## 1. The R--W metric violates the velocity addition ruler of special relativity

The modern stander cosmology is based on the Einstein's equation of gravitational field and the Robertson—Walker metric. According to the principle of cosmology, when the distribution of material is uniform and isotropic, the universal space-time can be described by the Robertson—Walker metric

$$ds^2 = c^2 dt^2 - R^2(t)\left( \frac{d\bar{r}^2}{1-\kappa\bar{r}^2} + \bar{r}^2 d\theta^2 + \bar{r}^2 \sin^2\theta \, d\varphi^2 \right) \qquad (6.1)$$

In the formula, $R(t)$ is the universal scalar factor, $\kappa$ is the curvature constant and $\bar{r}$ is the following coordinate. According to the principle of general relativity, we can use the resting reference frame of the earth to discuss the problem of cosmology. Suppose that light moves along the direction of radius in the flat universe, we have $ds = 0$, $\kappa = 0$ and $\theta = 0$. So according to (1), the motion of light in the expansive universe satisfies following quation

$$\frac{d\bar{r}}{dt} = \pm \frac{c}{R(t)} \tag{6.2}$$

When light moves along the direction at which $\bar{r}$ decrease, negative sign is taken, while light moves along the direction at which $\bar{r}$ increase, positive sign is taken. We can let $r(t) = R(t)\bar{r}$. In the flat universe, we can regard $r(t)$ as the common coordinates of both the light and celestial body. For the celestial body in the expensive universe, $\bar{r}$ does not change with time. But for the light moving in the expansive universe, $\bar{r}$ changes with time as described by (6.2). Suppose that there is an illuminant celestial body located at the point $r(t)$, relative to the resting earth reference frame, the velocity of celestial body is

$$V(t) = \frac{dr(t)}{dt} = \dot{R}(t)\bar{r} \tag{6.3}$$

So for the observers on the earth reference frame, the velocity of light emitted by the celestial body in the expansive universe is

$$V_c(t) = \frac{dr(t)}{dt} = \bar{r}\frac{d}{dt}R(t) + R(t)\frac{d\bar{r}}{dt} = \dot{R}(t)\bar{r} \pm c = V(t) \pm c \tag{6.4}$$

When the celestial body moves towards the observers, the positive sign is taken. While the celestial body moves away from the observers, the negative sign is taken. The formula above has nothing to do with whether or not there exists gravitational field in space and whether or not the celestial body has acceleration. When speed $V(t)$ is equal to a constant, or the universe expands in a uniform speed, it still holds. The formula indicates that when light moves in the expansive universe, its velocity satisfies the Galileo addition ruler of classical mechanics, not the Einstein's addition ruler of special relativity. This result obviously violates the principle of invariance of light's velocity and contradicts with physical experiments and astronomical observations. On the other hand, when we discuss the problems of cosmology at present, light's speed is always regarded as a constant without considering the existence of the formula (4). This is also inconsistent. If the Robertson—Walker metric is used in cosmology, we should use (4) to represent the velocity of light emitted by the illuminant celestial body in the expansive universe. But if (4) was used, many things would be changed.

As for the curve universe with $\kappa \neq 0$, the definition of time is still uniform in the whole space. Let $\theta = 0$ in (1), we have light's motion equation in the expensive universe

$$\frac{d\bar{r}}{dt} = \pm \frac{c\sqrt{1 - \kappa \bar{r}^2}}{R(t)} \tag{6.5}$$

According to the current theory, we can define the proper distance

$$r(t) = R(t)\int_0^{\bar{r}_1} \frac{d\bar{r}}{\sqrt{1 - \kappa \bar{r}^2}} = R(t)l(\bar{r}_1) \tag{6.6}$$

In which $l(\bar{r}_1) = \kappa^{-1/2} \sin n^{-1}(\kappa^{1/2}\bar{r}_1)$ corresponds to $\bar{r}$ in the flat universe. For the celestial body moving in the expansive universe, $l(\bar{r})$ also does not change with time. The celestial body's velocity relative to the earth reference frame is $V(t) = \dot{R}(t)l(\bar{r}_1)$. When light emitted by the celestial body moves in the curving expansive universal, we still have

$$V_c(t) = \frac{dr(t)}{dt} = \dot{R}(t)\int_0^{\bar{r}_1} \frac{d\bar{r}}{\sqrt{1 - \kappa \bar{r}^2}} + R(t)\left(\frac{d}{d\bar{r}}\int_0^{r_1} \frac{d\bar{r}}{\sqrt{1 - \kappa \bar{r}^2}}\right)\frac{d\bar{r}}{dt}$$

$$= \dot{R}(t)l(\bar{r}_1) + \frac{R(t)}{\sqrt{1-\kappa\bar{r}^2}}\frac{d\bar{r}}{dt} = V(t) \pm c \tag{6.7}$$

It is obvious that the light's velocity still satisfies the Galileo's addition ruler, in stead of the Einstein's addition ruler. Therefore, the Robertson---Walker metric can not be used as the basic frame of space-time to describe the expensive universe. The Friedmann equation of cosmology based on the Robertson---Walker metric and the Einstein's equation of gravitational field would also becomes unsuitable. Unfortunately, this fact has been neglected by physical circle for a long time, so that a serious aftermath is caused. Of course, we can let $R(t) = 1$ in (6.1) so that the Robertson---Walker metric becomes

$$ds^2 = c^2 dt^2 - \frac{dr^2}{1-\kappa r^2} - r^2 d\theta^2 - r^2 \sin^2\theta \, d\varphi^2 \tag{6.8}$$

In this way, though the contradiction with special relativity can be avoided, what we obtain is not the original Friedmann equation of cosmology again by connecting (8) with the Einstein's equation of gravitational field. Because (6.8) represents a static metric, it would be a problem whether or not it can be suitable to describe the expansive universe. Besides, there exists a deeper problem. The observation results of WMAP[1] indicate that our universe is almost flat. According to general relativity, the curve of space-time is caused by gravitational field. If space-time is flat, it means that there is no gravitational field. Let $\kappa = 0$ in (6.8), we get the metric of flat space-time. If we use this metric as the space-time frame of cosmology, it means that there is no gravitational interaction between materials in the universe. For the expensive universal, all celestial bodies would move in uniform velocities without acceleration or deceleration. Of course, this is impossible. But if we take the other form's metrics of curving space-time, the principle of cosmology would be violated. In our universe, the uniform and isotropic distribution of material is an observational fact. There exists a basic contradiction here. The solution of this fundamental contradiction may help us to understand the essence of our universal space-time and gravitation more deeply.

Meanwhile, we can also transformation the Robertson---Walker metric into other forms in light of the principle of general relativity. For example, we can introduce transformation[2] $\bar{r} = r/(1+\kappa r^2/4)$ and $\bar{R}(t) = R(t)/(1+\kappa r^2/4)$. By defining conformal time $d\eta = dt/\bar{R}(t)$, we can transform formula (1) into

$$ds^2 = \bar{R}^2(t)\left(c^2 d\eta^2 - dr^2 - r^2 d\theta^2 - r^2 \sin^2\theta \, d\varphi^2\right) \tag{6.9}$$

The contradiction with special relativity can also be avoided. But in this case, we obtain the new Friedmann equation after the Einstein's equation of gravitational field is considered. By using new metric and the new Friedmann equation, all relative problems would be re-calculated. We can not certain whether or not it is proper. In general, the results would be changed. Besides the Robertson---Walker metrics, the other cosmological metrics which is similar to the Robertson---Walker metric have the same problem. For example, for the de Sitter metric, we have

$$ds^2 = c^2 dt^2 - e^{2Ht}\left(d\bar{r}^2 + \bar{r}^2 d\theta^2 + \bar{r}^2 \sin^2\theta \, d\varphi^2\right) \tag{6.10}$$

Let $R(t) = exp(2Ht)$, we get $V(t) = 2H\bar{r}\, exp(2Ht)$ and $V_c(t) = V(t) \pm c$. The result is the same. So the any R-W type metric based on the universal scalar factor to describe the universal expansion would face a same problem of violating the velocity addition ruler of special relativity.

Therefore, the standard theory of modern cosmology is facing real crisis. Some established viewpoints and conclusions in the stander cosmology should be re-surveyed. The most direct problems are dark energy and the accelerating expansion of the universe which are discussed below. The current formula used to calculate the resfhift of cosmology is based on the Robertson---Walker metric directly. Suppose that a celestial body located at the coordinate point $r_1$ emits light with frequency $v_1$ and period $\Delta t_1$ at time $t_1$. When the light arrives at the original point of reference frame at time $t_0 > t_1$, the observers located at the original point find that the light's frequency and period become $v_0$ and $\Delta t_0$. Based on formula (6.5), we can deduce the relation [2]

$$\frac{\Delta t_0}{R(t_0)} = \frac{\Delta t_1}{R(t_1)} \tag{6.11}$$

On the other hand, the frequency of light can not be measured directly. What observed in the experiments is wave length actually. Suppose that light's speed is also unchanged in the expansive universe, the definition of resdshift should be

$$1 + Z = \frac{\lambda_0}{\lambda_1} = \frac{v_1}{v_0} = \frac{\Delta t_0}{\Delta t_1} = \frac{R(t_0)}{R(t_1)} \tag{6.12}$$

By defining luminosity distance $d_L = R(t)r(1+Z)$ and using the Friedmann equation of cosmology, we can deduce the relation between $d_L$ and $Z$ [3]

$$\frac{H_0}{c} d_L = \frac{1+Z}{|\Omega_k|^{1/2}} sinn |\Omega_k|^{1/2} \int_0^z \frac{dz'}{\sqrt{\Omega_k (1+z')^2 + \Omega_{m0}(1+z')^3 + \Omega_\lambda}} \tag{6.13}$$

Here $\Omega_{m0}$ is the universal material density at present time $t_0$, and $\Omega_\lambda$ is the so-called dark energy density corresponding to vacuum and universal constant. When $\Omega_\lambda = 0$, (13) can be simplified as [4]

$$\frac{H_0}{c} d_L = \frac{2\Omega_{m0} Z + (2\Omega_{m0} - 4)(\sqrt{\Omega_{m0} Z + 1} - 1)}{\Omega_{m0}^2} \tag{6.14}$$

By the fit between the formula (13) and the practical observations of the high red-shift Ia type supernovae, the results of $\Omega_{m0} = 0.3$ and $\Omega_\lambda = 0.7$ as well the accelerating expansion of the universal are deduced [5]. But if the Robertson—Walker metric is unsuitable, the formula (6.12) can not be used. All these deductions about dark energy and the accelerating expansion of the universe would be given up.

On the other hand, according to (6.4) and (6.7), we would have $v_1 \lambda_1 = c(t_1) = c(1 + V_1/c)$. The definition of shown in (6.12) become

$$1 + Z = \frac{\lambda_0}{\lambda_1} = \frac{v_1}{v_0} = \frac{1}{1 + V_1(t_1)/c} \frac{\Delta t_0}{\Delta t_1} = \frac{1}{1 + \dot{R}(t_1)\bar{r}/c} \frac{R(t_0)}{R(t_1)} \tag{6.15}$$

By using the Friedmann equation of cosmology, for the flat universe with $k = 0$, we can get

$$1 + Z = \frac{1}{1 + H(t_1)R(t_1)\bar{r}/c} \frac{R(t_0)}{R(t_1)} \tag{6.16}$$

For the high redshift supernovae, the ratio $H(t_1)R(t_1)\bar{r}/c \sim V_1/c$ can not be neglected. So from the formula above, we can not get the formula (6.13) as well as the results $\Omega_{m0} = 0.3$ and $\Omega_\lambda = 0.7$, though the formula (6.16) is also unacceptable for it violates the velocity addition ruler of special relativity.

From other hands, we can also prove that the formula (6.12) is unsuitable as the foundation to calculate the high redshift of supernovae. Suppose that there is no gravitational field in space. The illuminant celestial body moves away from the observers in a uniform velocity and arrives at the point $r_0 > r_1$ at time $t_0 > t_1$. If (6.12) is correct, we would obtain the Doppler redshift from the formula. However, this is impossible. Because the light emitted by the celestial body arrives at the original point from the point $r_1$ during the period $\Delta t = t_0 - t_1$, we have $r_1 / \Delta t = c$. According to (6.12), we get

$$Z = \frac{\bar{r}R(t_0) - \bar{r}R(t_1)}{\bar{r}R(t_1)} = \frac{r_0 - r_1}{r_1} = \frac{V\Delta t}{r_1} = \frac{V}{c} \tag{6.17}$$

It seems that we obtain the relation of the Hubble redshift—velocity, but this is only an approximate formula. It is tenable only when $V \ll c$. We can only use it to describe low redshift. We can not use it to describe the high redshift of supernovae. When there is no gravitational field, the formula of redshift caused by the Doppler effect is

$$Z = \sqrt{\frac{1 + V/c}{1 - V/c}} - 1 \tag{6.18}$$

Because the calculation of (6.17) is strict when there is no gravitation, if (6.12) is correct, we should get (6.18) in stead of (6.17) from it. In fact, when there is gravitation in space, for the situation with $k = 0$, $\Omega_{m0} = 1$ and $\Omega_\lambda = 0$, we can deduce from (6.14)

$$Z = \left(1 - \frac{H_0 r}{2c}\right)^{-2} - 1 \to \left(1 - \frac{V}{2c}\right)^{-2} - 1 \tag{6.19}$$

It is obvious that the form of (6.19) is completely different from (6.18). So (6.14) can not describe the real Doppler redshift actually.

We can also prove that (6.14) can not be used to describe gravitational redshift. In fact, if there is no gravitational field in space, the gravitational redshift should be zero. The gravitational redshift is only relative to acceleration, having nothing to do with velocity. But as we known that the redshift described by (6.17) is non-zero when the celestial body moves in a uniform velocity without acceleration. As we known in physics, there are only ways to cause redshift, i.e., the Doppler and gravitational effects. The cosmological redshift is generally considered to be caused by the Doppler effect at present. The formula (6.13) based on (6.12) describes neither the real Doppler redshift nor gravitational redshift. Its origin is distrustful.

In fact, the idea that weird dark energy governs our universe has given us a hint that something may be wrong in our theory of cosmology. Because the Robertson—Walker metric is unsuitable as the basic frame of space-time to describe the expensive universe, we would face a task to reestablish the theoretical foundation of cosmology and re-survey the conclusions in the current theory of cosmology. The results would produce important influence on the foundation of modern physics.

## 3. The reference frame to describe the universal expansion

In order to describe the universal expansion simply and properly, we need to establish a proper reference frame. Thought special and general relativities deny the existence of the absolutely resting reference frame, the big-bang cosmology actually implicates the existence of this kind of special reference frame. In light of the current viewpoint, the universe originated from a primordial big-bang. The big-bang means the existence of an original point. We can take this point as the original point to establish a static reference frame, called as the universal big-bang reference frame. In the expansive process of the universal, all celestial bodies and material are considered to move relative to this reference frame. In fact, in the 1960's, astronomers found the spatial anisotropy of microwave background radiation. If the reference frame in which microwave background radiation was isotropic was taken as the resting reference frame in the process of the universal expansion, observations showed that the earth was moving in a speed $300 Km \cdot s^{-1}$ towards to the directions of right ascension $1^h.5 \pm 0^h.4$ and declination $0.^02 \pm 7^0$ [5]. In 1999, the anisotropy detector of microwave background radiation (WMAP) found anisotropy in a higher precision [6]. In 2002, physicists found the anisotropy of radio waves eradiated by radio galaxy in the direction of the earth's motion by using array radio telescopes (VLA) [7]. This kind of anisotropy can also be explained by the Doppler effect of the earth's motion. So by means of these measurements of spatial anisotropy of microwave background radiation, we can already determine the orientation of the universal big-bang reference and the motions of other celestial bodies relative to it. Only by the restriction of the special and general principles of relativity, we now have no enough courage to admit it.

In fact, when we sue the Einstein's theory of gravitation to research the problem of the universal expansion, we also need a reference frame based on it the motion can be established. At present, we choose the reference frame of the sun system or the Milky Way galaxy as the resting reference frame, in which the resting earth is considered as the original point or the center of the universe. But the observations have shown that the earth is not the center of the earth. But we prefer to choose the universal big-bang reference frame as the reference frame to descrier the universal expansion to the earth reference frame, no matter whether or not the absolutely resting reference frame or the center of the universe exists. In the following discussion, by the consideration of logical rationality, simplification and applicability, we study the problem of cosmology based on the universal big-bang reference frame.

## 4. Velocity and Acceleration of the Universal Expansion

It seems to be a common idea that only general relativity could provide a proper foundation for the discussion of cosmology at present. However, it was pointed by E. A. Milne in 1943 that the Newtonian formula of gravity could also be used to describe the expansion of the universe [22]. The motion equation of the universal expansion deduced from the Newtonian theory was similar to that from general relativity, except there is no the item containing cosmic constant. It is proved below that when the formula (26) is used to describe the universal expansion, the revised Hubble formula can be deduced and the departure from the linear relation of distance---red-shift observed in the high red-shift type Ia supernovae can be explained well. Because there is no repulsive force in this theory, the hypotheses of the universal accelerating expansion and dark energy become unnecessary.

Suppose there is a medium sphere with radius $R$ and density $\rho_0$. The static mass of sphere is $M_0 = 4\pi\rho_0 R^3 / 3$. According to the Newtonian theory, the gravitation force acted on an object with static mass $m_0$ located at the point $r$ outside or inside the sphere are individually [23]

$$F = -\frac{GM_0 m_0}{r^2} \quad r \geq R \qquad \text{or} \qquad F = -\frac{GM_{0r} m_0}{r^2} \quad r \leq R \qquad (6.20)$$

Here $M_{0r} = 4\pi\rho_0 r^3 / 3$ is the static mass of sphere with radius $r$. The formulas indicate that when mass $m_0$ is located outsider the sphere with $r > R$, the gravitation acted on it is equal to that when the spherical mass is centralized at the spherical center. When mass $m_0$ is located inside the sphere with $r < R$, the gravitation acted on it is only relative to the spherical mass $M_{0r}$, having nothing to do with the mass distributed outside the radius $r$. It is obvious that when $R \to \infty$, the conclusion above is still tenable. We would show below that the conclusion also holds for the revised Newtonian formula when object's angle velocity $\vec{L}=0$.

Suppose that the universe expands along the radius direction. In the expansion process, the angle momentum $\vec{L}$ of object is equal to zero. We discuss the problems by the method of stage by stage approximation. Suppose again that an object is located at the point $r$ in the big-bang reference frame, its velocity $V_r$ satisfies (4.22) approximately at first. We consider a spherical shell with radius $R$ of which the center is just at the original point of the big-bang reference frame. Let $\sigma$ be the mass density of spherical shell. Meanwhile, there is an object located at the point $r > R$ with static mass $m_0$ and velocity $\vec{V}_r$ along the radius. We calculate the gravitation that the spherical shell acts on the object. Because $R$ is a constant at a certain moment, the formula (6.20) is still effective as long as gravitational static masses are substituted by gravitational moving masses. Because the static mass of spherical shell is $M_{0\sigma} = 4\pi\sigma R^2$, according to (5.2), we have the gravitation that the spherical shell acts on the object

$$F_\sigma = -\frac{Gm_0 4\pi\sigma R^2}{r^2}\sqrt{1-\frac{V_R^2}{c^2}}\sqrt{1-\frac{V_r^2}{c^2}}$$

$$= -\frac{Gm_0 4\pi\sigma R^2}{r^2}\frac{\sqrt{1-V_r^2/c^2}}{\sqrt{1+\alpha_R/R}} = -\frac{GM_\sigma m}{r^2} \qquad (6.21)$$

$$M_\sigma = \frac{4\pi\sigma R^2}{\sqrt{1+\alpha_R/R}} = 4\pi\sigma R^2\sqrt{1-\frac{V_R^2}{c^2}} \qquad m = m_0\sqrt{1-\frac{V_r^2}{c^2}} \qquad (6.22)$$

Here $\alpha_R = 2GM_{0R}/c^2 = 8G\pi\rho_0 R^3/(3c^2)$, $\alpha_r = 2GM_{0r}/c^2 = 8G\pi\rho_0 r^3/(3c^2)$. $M_{0R}$ and $M_{0r}$ are the static masses of spheres with the same densities $\rho_0$ but different radius $R$ and $r$. Because $R$ is a constant for a sphere shell, the formula (6.21) represents the gravitation acted on the moving object and caused by the spherical shell when its gravitational moving mass is considered to centralize at the spherical center. Let $b = 8G\pi\rho_0/(3c^2)$, $x = \alpha_R/R = bR^2$. By substituting both into (6.21) and taking the integral over $R$, we get the total gravitation that the expensive sphere with radius $R = r$ acts on an object with static mass $m_0$ and velocity $V_r$ located on the spherical surface

$$F = -\frac{Gm_0 4\pi\sigma}{r^2}\sqrt{1-\frac{V_r^2}{c^2}}\int_0^r \frac{R^2 dR}{\sqrt{1+bR^2}}$$

$$= -\frac{GM_{0r}m_0}{r^2}\frac{3}{2}\frac{\sqrt{x+x^2}-ln(\sqrt{x}+\sqrt{1+x})}{x^{3/2}}\sqrt{1-\frac{V_r^2}{c^2}} \qquad (6.23)$$

On the other hand, we have

$$F = \frac{d}{dt}\frac{m_0 V_r}{\sqrt{1-V_r^2/c^2}} = \frac{m_0\ddot{r}}{(1-V_r^2/c^2)^{3/2}} \qquad (6.24)$$

From both formulas above, we obtain the acceleration of an object on the spherical surface in the processes of the universal expansion

$$\ddot{r} = -\frac{GM_{0r}}{r^2}\left(1-\frac{V_r^2}{c^2}\right)^2 = -\frac{GM_0}{r^2}Q_1(x) = -\frac{GM}{r^2} \qquad (6.25)$$

In which $M = M_0 Q_1(x)$. The formula indicates that for the expansive universe, we can use equivalent mass $M$ to substitute the static mass $M_0$ in the Newtonian theory. We have

$$Q_1(x) = \frac{3}{2} \frac{\sqrt{x+x^2} - \ln(\sqrt{x}+\sqrt{1+x})}{x^{3/2}(1+x)^2} \tag{6.26}$$

By the same consideration, it is easy to understand that when $R > r$ and $\vec{L} = 0$, the resultant gravitation caused by the spherical shell, acted on an object which is located inside the spherical shell, is also zero. But it is unnecessary for us to discuss any more here.

So when we discuss the problems of cosmology based on the big-bang reference frame, the gravitation acted on an object which is located at point $r$ is only relative to the mass contained in the spherical shell with radius $r$, having nothing to do with the total mass of the universe, no mater whether the universe is finite or infinite. Suppose that the total mass contained in the spherical shell is $M_0$ it is enough for us only to consider the gravitation caused by $M_0$, acted on the object located on the surface of sphere. Because mass $M_0$ is finite, when the spherical radius $r \to \infty$ in the process of the universal expansion, we have $x \to 0$. When the spherical radius $r \to 0$ in the process of the universal contraction, we have $x \to \infty$. It is easy to prove the following limitations

$$\lim_{x \to 0} Q_1(x) = \lim_{x \to 0} \frac{3}{\sqrt{1+x}(1+x)^2(3+2x)} = 1 \tag{6.27}$$

$$\lim_{x \to \infty} Q_1(x) = \lim_{x \to \infty} \frac{3}{\sqrt{1+x}(1+x)^2(3+2x)} = 0 \tag{6.28}$$

By the numerical calculation, it can be known that we always have $Q(x) > 0$ and $\ddot{r} \leq 0$ within the region $0 < x < 1$. So the expansive speed of the universe is always decreased, that is to say, no the universal accelerating expansion actually. In order to know object's velocity in the expansive process, by considering relation $\ddot{r} = dV_r/dt = V_r dV_r/dr$ and taking the integral of (6.37), we have

$$V_r^2 - V_{r_0}^2 = -\int_{r_0}^{r} \frac{GM_0}{r^2} Q_1(x) dr = 3GM_0 \int_{x_0}^{x} \frac{\sqrt{x'+x'^2} - \ln(\sqrt{x'}+\sqrt{1+x'})}{x'^{3/2}(1+x')^2} dx' \tag{6.29}$$

It is difficult to complete the integral. But we can do approximate calculation. We have following developing formulas within the region $-1 < x < 1$

$$\frac{1}{\sqrt{1+x}} = 1 - 0.50x + 0.38x^2 - 0.31x^3 + 0.27x^4 - 0.25x^5 + 0.23x^6 \cdots$$

(6.30)

$$\frac{1}{(1+x)^2} = 1 - 2x + 3x^2 - 4x^3 + 5x^4 - 6x^5 + 7x^6 \cdots \tag{6.31}$$

By substituting (6.20) into (6.23) and considering (6.21) and (6.24), we get at last

$$\ddot{r} = -\frac{GM_0}{r^2}\left(1 - \frac{2.30\alpha}{r} + \frac{3.76\alpha^2}{r^2} - \frac{5.32\alpha^3}{r^3} + \frac{6.95\alpha^4}{r^4} - \frac{8.64\alpha^5}{r^5} + \frac{9.51\alpha^6}{r^6} \cdots\right) \tag{6.32}$$

$$V_r^2 = \frac{2GM_0}{r}\left(1 - \frac{1.15\alpha}{r} + \frac{1.25\alpha^2}{r^2} - \frac{1.33\alpha^3}{r^3} + \frac{1.39\alpha^4}{r^4} - \frac{1.44\alpha^5}{r^5} + \frac{1.46\alpha^5}{r^5} \cdots\right) + A \tag{6.33}$$

Let $V_r = 0$ when $r \to \infty$, we have integral constant $A = 0$. Then considering (6.33) as the more accurate speed of an object located on the expansive spherical surface, substituting it into (6.23) and doing the second time of calculation in light of the same procedure, we get the second approximate results of the acceleration and velocity. Let $V_r \to V$, we have at last

$$\ddot{r} = -\frac{GM_0}{r^2}Q_1(r) = -\frac{GM}{r^2} \qquad\qquad V^2 = \frac{2GM_0}{r}Q_2(r) \qquad (6.34)$$

$$Q_1(r) = 1 - \frac{2.30\alpha}{r} + \frac{4.09\alpha^2}{r^2} - \frac{6.62\alpha^3}{r^3} + \frac{8.74\alpha^4}{r^4} - \frac{11.55\alpha^5}{r^5} + \frac{13.86\alpha^5}{r^5} \cdots \qquad (6.35)$$

$$Q_2(r) = 1 - \frac{1.15\alpha}{r} + \frac{1.36\alpha^2}{r^2} - \frac{1.57\alpha^3}{r^3} + \frac{1.75\alpha^4}{r^4} - \frac{1.93\alpha^5}{r^5} + \frac{2.13\alpha^6}{r^5} \cdots \qquad (6.36)$$

It can be seen that the differences only appear on the third and later items. Comparing with the Newtonian theory, it is known that in the expansive process of the universe, the speed and acceleration caused by gravitation would become small. It is equal to let $M_0 \to M = M_0 Q_1(r)$ in the motion equation, in which $M$ is not the moving mass of gravitation, but can be considered as the equivalent mass in the universal expansive process.

### 3. The cosmological redshift and the Hubble diagram of

From (6.33) we get the Hubble law of cosmological redshift $V = Hr$ in which

$$H = H_1 \sqrt{1 - 1.15\left(\frac{H_1 r}{c}\right)^2 + 1.36\left(\frac{H_1 r}{c}\right)^4 - 1.57\left(\frac{H_1 r}{c}\right)^6 + 1.75\left(\frac{H_1 r}{c}\right)^8 \cdots} \qquad (6.37)$$

Here $H_1 = \sqrt{8\pi G\rho_1/3}$ is the Hubble constant. When $\alpha/r = (H_1 r/c)^2 << 1$, we get the Hubble formula from the formula above

$$Z_d = \frac{V_1}{c} = \frac{H_1}{c}r \qquad (6.38)$$

It is noted that according to the current definition, the Hubble constant $H_0 = \sqrt{8\pi G\rho_c/3}$. Here $\rho_c$ is the critical material density, in spite of the material density $\rho$. By the astronomy observation, we can take $H_1 = 65 Km \cdot s^{-1} \cdot Mpc^{-1} = 2.1 \times 10^{-18} s^{-1}$ at present and get $H_1/c = 7 \times 10^{-27} m^{-1}$. So (6.38) is only suitable for the situation with $r < 10^{25} m$.

Thought the Hubble formula (6.38) is defined in the big-bang reference frame, it is easy to prove that the formula is also suitable for the observation on the earth reference frame. As shown in Fig.6.1, suppose that $r_1$ and $\vec{V}_1$ are the distance and velocity of the earth relative to the big-bang reference frame, $r_2$ and $\vec{V}_2$ are the distance and velocity of a certain celestial body relative to the big-bang reference frame located at arbitrary direction $\theta$, $r$ and $\vec{V}$ are the distance and velocity of the celestial body relative to the earth, we have $V_1 = H_0 r_1$, $V_2 = H_0 r_2$ and obtain

$$V = \sqrt{V_1^2 + V_2^2 - 2V_1 V_2 \cos\theta} = H_1\sqrt{r_1^2 + r_2^2 - 2r_1 r_2 \cos\theta} = H_1 r \qquad (6.39)$$

So the Hubble distance--redshift linear relation between the earth and celestial bodies at arbitrary direction still holds when the revised items are neglected. When the observers on the earth measures the distant celestial bodies with red-shift $Z >> 10^{-3}$, the distance between the earth and the original point of the big-bang reference frame can be neglected, so that the red-shift observed on the earth can be considered as that observed at the big-bang reference frame. We discuss the problems of the high red-shift type Ia supernova in this way below.

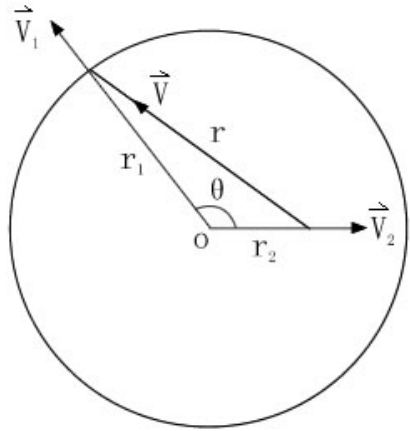

**Fig. 6.1 The Hubble redshift --distance relation between the earth
and the celestial bodies at arbitrary space directions**

It should be noted that in the formula above, the Hubble constant $H_1$ and material density $\rho_1$ take the values of past time $t_1$ while the celestial body is located at the position $r_1$. In the practical observations, the observed redshift is also the value of past time $t_1$ while the celestial body is located at the position $r_1$. At the present time $t_2$, the celestial body has moved to the position $r_2$. For the cosmological process, the difference between $t_1$ and $t_2$ can reaches the magnetic order of the universal age, the difference between $r_1$ and $r_2$ is also great. Our purpose is to determine the Hubble constant $H_0$ and the universal material density $\rho_0$ of the present time by comparing the practical observations and the theoretical results. In the following discussion, we firstly use the Hubble constant $H_1$ and the universal material density $\rho_1$ of the present time to represent the redshift of cosmology then transform it into the value represented by the Hubble constant and the universal material density of the present time.

In the formula 6.34 , we suppose that there only exists gravitational interaction between the material in the universe. Meanwhile, the velocities $V \to 0$ on the surface of medium sphere when $r \to \infty$. This is actually the model of the infinite expansion of the universe. The practical situation would be that when the material density of the universe was great, the other form's interaction could not be neglected. Under the condition of high temperature and high pressure in the early universe, the electromagnetic interaction between charged particles as well as weak and strong interaction should be considered. Under the higher energy condition in the earlier universe, there may exist unknown interaction. So in order to describe the early expansion of the universal rationally, the other interaction should be considered. At present, the universe is considered to be born from a singularity, or the universal material can be compressed into a point by gravitation. This is unreliable and unbelievable. In the real word, infinite density is impossible. The physical history tell us that the advancement of physics always concomitances with the elimination of infinities. Foe example, the infinite elimination of ultraviolet radiation accompanies the birth of quantum theory. The infinite elimination the propagation speed of interaction accompanies the birth of special relativity. The infinite eliminations in high order perturbation process declare the success of quantum electrodynamics and the unity theory of electro-weak interaction. The non-normalization and singularity of the Einstein's theory of gravitation actually indicates that there exist some foundational problems in the theory. In order to avoid the infinite density caused by gravitation, we suppose that there exists a certain mechanism or an interaction so that a sphere with static mass $M_0$ can only be compressed into that with radius $r_0$. In this way, the motion equation of the universal expansion can be written as

$$\ddot{r}_1 = -\frac{GM_0}{r_1^2}Q_1(r_1) + F(r_1) \tag{6.40}$$

In which $F(r_1)$ is the sum of all other interaction forces besides gravitation with the following form

$$F(r_1) = \frac{1}{2}A(r_1)\delta(r_1 - r_0) \tag{6.41}$$

It indicates that there exists an infinite potential barrier at point $r_0$ caused by all other interaction forces, so that a spherical surface with radius $r_1$ can only be shrunk into that with radius $r_0$. Meanwhile, by this force's action, the spherical surface would expanse with a positive speed, so that contracting process would becomes expansive process. If let $r_0 \to 0$, it becomes the universe bursting out from a singularity. Similar to (6.24), substituting (6.41) into (6.40) and considering (6.34), then by taking the integral, we get

$$V_1^2 = \frac{2GM_0}{r_1}Q_2(r_1) + A(r_0) + C \tag{6.42}$$

By considering the fact that at the initial time when $r = r_0$, the second item (non-gravitational interaction) in (6.40) may be far greater than the first item (gravitational interaction), we can chose the initial condition so that to let $V_1^2 = A(r_0)$, or we can let $C = -2GM_0/r_0$. In this way, we have from the formula above

$$V_1^2 = \frac{2GM_0}{r_1}Q_2(r_1) + A(r_0) - \frac{2GM_0}{r_0}Q_2(r_0) \tag{6.43}$$

Then let

$$A(r_0) = 8G\pi\rho_0 r_0^2 \eta(r_0) \tag{6.44}$$

In which $\eta(r_0)$ is an unknown function. In this way, the second two items in (6.42) can be written as

$$A(r_0, \eta) - \frac{2GM_0}{r_0}Q_2(r_0) = \frac{8G\pi\rho_0 r_0^2}{3}[\eta(r_0) - Q_2(r_0)] \tag{6.45}$$

Due to the law of mass conservation in the universal expansion process, we have

$$M_0 = \frac{4\pi\rho_1 r_1^3}{3} = \frac{4\pi\rho_0 r_0^3}{3} \tag{6.46}$$

The formula (6.43) ca be written as

$$V_1 = r_1\sqrt{\frac{8\pi G\rho_1}{3}}\sqrt{Q_2(r_1) + \frac{r_1}{r_0}[\eta(r_0) - Q_2(r_0)]} \tag{6.47}$$

We define the Hubble constant

$$H = \sqrt{\frac{8\pi G\rho}{3}} \tag{6.48}$$

For the uniform expansion of the universe, we always have $r_1/r_0 = b = $ constant, here $b$ is the multiples of the universal expansion, so the formula (6.47 becomes

$$V_1 = H_1 r_1 \sqrt{Q_2(r_1) + b[\eta(r_0) - Q_2(r_0)]} = H_1 r_1 \sqrt{Q_2(r_1) + \kappa(b, r_0)} \tag{6.49}$$

In which

$$\kappa(b, r_0) = b[\eta(r_0) - Q_2(r_0)] \tag{6.50}$$

$\kappa(b, r_0)$ is relative to $r_0$. For a certain celestial body, $\kappa(b, r_0)$ is a constant in the expansive process of the universe. But for different celestial bodies, $\kappa(b, r_0)$ is different for their initial positions $r_0$ are different. Only when all material sprayed out from a sing singularity, $\kappa(b, r_0)$ would be a constant. In general situations, we have $\kappa(b, r_0) > 0$ and $\kappa(b, r_0) \leq 0$. When $\kappa(b, r_0) > 0$ and $r_1 \to \infty$, according to (6.43), we get

$$V^2(r_1 \to \infty) \to A(r_0) - \frac{2GM_0}{r_0} Q_2(r_0) > 0 \tag{6.51}$$

It means that our universe would keep on expansion all along until the material density became infinitely thin. For the situation with $k(r_0) < 0$, suppose when $r = R$ we have $V(R) = 0$, form (6.43) we get

$$\frac{2GM_0}{R} Q_2(R) = -\left[A(r_0) - \frac{2GM_0}{r_0} Q_2(r_0)\right] > 0 \tag{6.52}$$

It means that the universal expansion would stop. After that, it began to re-contract. Our universe would be a circulatory universe to expanse and contract. On the other hand, for the celestial body which is near the earth, we have $\alpha / r \ll 1$ and $Q_2(r) \approx 1$, so that we have

$$Z_d = \frac{V_1}{c} = \frac{H_1 r_1}{c} \sqrt{1 + \kappa(b, r_0)} \tag{6.53}$$

Because for different celestial bodies at different positions at present, their initial positions are different, even under the condition $\alpha / r \ll 1$, the Hubble constant is not a real constant. This may be the reason why we can not measure the Hubble constant accurately up to now.

If the change of light's speed in gravitational field is considered, the Doppler redshift shown in (4.116) should be written as

$$Z_d = \left(1 + \frac{\alpha}{r_1}\right) \sqrt{\frac{1 + V_1/c}{1 - V_1/c}} - 1 = \left(1 + \frac{\alpha}{r_1}\right) \sqrt{\frac{1 + H_1 r_1 \sqrt{Q_2(r_1) + \kappa(b, \eta, r_0)}/c}{1 - H_1 r_1 \sqrt{Q_2(r_1) + \kappa(b, \eta, r_0)}/c}} - 1 \tag{6.54}$$

By considering the influence of the universal expansion on the mass of material, according to (6.46), let $M_0 \to M_0 Q_1(x)$ in $\alpha$, (6.54) should be written

$$Z_d = \left(1 + \frac{H_1^2 r_1^2 Q_1(r_1)}{c^2}\right) \sqrt{\frac{1 + H_1 r_1 \sqrt{Q_2(r_1) + \kappa(b, \eta, r_0)}/c}{1 - H_1 r_1 \sqrt{Q_2(r_1) + \kappa(b, \eta, r_0)}/c}} - 1 \tag{6.55}$$

Then let's discuss how to transform the Hubble constant of past time into that of present time, so that we can compare the rhetorical results with experiments. Owing to the law of mass conservation, we have

$$H_1 = \sqrt{\frac{8\pi G \rho_1}{3}} = \sqrt{\frac{8\pi G \rho_2}{3}} \left(\frac{r_2}{r_1}\right)^{\frac{3}{2}} = H_0 \left(\frac{r_2}{r_1}\right)^{\frac{3}{2}} \tag{6.56}$$

Let $r_2^{3/2} / r_1^{3/2} = \varepsilon$, as long as we know the ratio $\varepsilon$, we obtain the relation $H_1 = \varepsilon H_0$. In the expansive process of the universe, when a celestial body moves to point $r_2$ from point $r_1$, the light emitted by the celestial body propagates to the observers located at the original point from point $r_1$. We now calculate expended time that light travels in the process. If the changing law of the universal material with time is unknown, this calculation is difficult. For simplification, we take the uniform values of material density during time $t_1$ and $t_2$ to let

$$\rho = \frac{\rho_2 + \rho_1}{2} = \frac{\rho_1}{2}\left(1 + \frac{r_1^3}{r_2^3}\right) = \frac{\rho_2}{2}\left(1 + \frac{r_2^3}{r_1^3}\right) \tag{6.57}$$

If the influence of the universal expansion on mass is not considered, in light of (4.70), we have

$$\Delta t = \frac{1}{c}\int_0^{r_1} \frac{dr}{1 + 8\pi G\rho \, r^2/(3c^2)} dr = \frac{\sqrt{2}}{H_1\sqrt{1 + r_1^3/r_2^3}} arctg \frac{H_1\sqrt{1 + r_1^3/r_2^3}}{\sqrt{2}c} r_1 \tag{6.58}$$

If the influence of the universal expansion on mass is considered, we can let $\rho_1 \to \rho_1 Q_1(r_1)$. By using (6.56), the formula can be written as

$$\Delta t = \frac{\sqrt{2}}{H_1\sqrt{Q_1(r_1)}\sqrt{1 + r_1^3/r_2^3}} arctg \frac{H_1 r_1 \sqrt{Q_1(r_1)}\sqrt{1 + r_1^3/r_2^3}}{\sqrt{2}c}$$

$$= \frac{\sqrt{2}}{H_0\sqrt{Q_1(r_1)}\sqrt{1 + \varepsilon^2}} arctg \frac{H_0 r_1 \sqrt{Q_1(r_1)}\sqrt{1 + \varepsilon^2}}{\sqrt{2}c} \tag{6.59}$$

During the same time, the celestial body moves to point $r_2$ from point $r_1$, according to (6.51), we get

$$\Delta t = \int_{r_1}^{r_2} \frac{\sqrt{r}\,dr}{\sqrt{2GM_0\sqrt{Q_2(r)} + \kappa}} \tag{6.60}$$

The formula can not be integrated. We have to take the uniform values to let $Q_2(r) \to [Q_2(r_1) + Q_2(r_2)]/2$. For a certain celestial body, $\kappa$ is a constant. Form (6.60) we obtain

$$\Delta t = \frac{2(1 - 1/\varepsilon)}{3H_0\sqrt{[Q_2(r_1) + Q_2(r_2)]/2 + \kappa}} \tag{6.61}$$

Form (6.59) and (6.61), we have at last

$$\frac{arctg H_0 r_1 \sqrt{Q_1(r_1)}\sqrt{1 + \varepsilon^2}/(\sqrt{2}c)}{\sqrt{Q_1(r_1)}\sqrt{1 + \varepsilon^2}} = \frac{\sqrt{2}(1 - 1/\varepsilon)}{3\sqrt{[Q_2(r_1) + Q_2(\varepsilon^{2/3} r_1)]/2 + \kappa}} \tag{6.62}$$

Under the condition $H_0 r_1 \sqrt{Q_1(r_1)}\sqrt{1 + \varepsilon^2}/(\sqrt{2}c) \ll 1$, the right side of the formula can be developing into the Taylor's series with

$$\frac{H_0 r_1}{c}\left\{1 - \frac{1}{3}\left[\frac{H_0 r_1 \sqrt{Q_1(r_1)(1 + \varepsilon^2)}}{\sqrt{2}c}\right]^2 + \frac{1}{5}\left[\frac{H_0 r_1 \sqrt{Q_1(r_1)(1 + \varepsilon^2)}}{\sqrt{2}c}\right]^4\right.$$

$$\left. - \frac{1}{7}\left[\frac{H_0 r_1 \sqrt{Q_1(r_1)(1 + \varepsilon^2)}}{\sqrt{2}c}\right]^6 + \cdots\right\} = \frac{2(1 - 1/\varepsilon)}{3\sqrt{[Q_2(r_1) + Q_2(\varepsilon^{2/3} r_1)]/2 + \kappa}} \tag{6.63}$$

It is noted that our purpose is to calculate $\varepsilon$ while $H_0 r_1/c$ is known. However, because $\kappa$ is unknown for different celestial bodies, we can not obtain the value $\varepsilon$ from formula above. So we can only use the method of successive approximation. The method is that for a certain value of $H_0 r_1 / c$, let $\kappa = \kappa_1$ and substitute it into (6.62) and calculate $\varepsilon_1$. Then let $H_1 = \varepsilon_1 H_0$ and put it into (6.55), get the redshift described by $H_0$. Comparing the result with practical observation, we can determine the new value of $\kappa_2$. Then we use new $\kappa_2$ in (6.62) to calculate $\varepsilon_2$ again. In this way, we can reach the consistent result with $\kappa_{n+1} = \kappa_n$ at last and $\kappa_n$ can be regarded as $\kappa$.

Using the results above in (6.56), we can write (6.55) as

$$Z_d = \left[1 + \frac{\varepsilon^2 H_0^2 r_1^2 Q_1(r_1)}{c^2}\right] \sqrt{\frac{1 + \varepsilon\, H_0 r_1 \sqrt{Q_2(r_1) + \kappa(b, r_0)}/c}{1 - \varepsilon\, H_0 r_1 \sqrt{Q_2(r_1) + \kappa(b, r_0)}/c}} - 1 \tag{6.64}$$

When $H_0 r_1 / c \ll 1$, we have $Q_1(r) \approx 1$, $Q_2(r) \approx 1$ and $\varepsilon \approx 1$, i.e., $Z_d \approx H_0 r_1 / c$.

Besides the Doppler redshift, we should consider the gravitational redshift in cosmological problems, for both exist really and with the same order of magnitude under the situation of high redshift. The gravitational redshift is discussed below. In this case, we can regard the universe as medium sphere in which mass is distributed uniformly. When light emitted by the celestial located on the surface of the sphere propagates in the sphere, the gravitation of mass would affect light's wave length and frequency. For simplification, we use (4.107) to calculate gravitational redshift approximately. Meanwhile, by considering the influence of the universal expansion on mass, in (4.107) we let

$$\frac{\alpha}{r_1} = \frac{2GM_0}{c^2 r_1} \rightarrow \frac{2GM_0 Q_1(r_1)}{c^2 r_1} = \frac{H_1^2 Q_1(r_1) r_1^2}{c^2} = \frac{\varepsilon^2 H_0^2 Q_1(r_1) r_1^2}{c^2} \tag{6.65}$$

In this way, the gravitational redshift can be described approximately

$$Z_g = \left[1 + \frac{\varepsilon^2 H_0^2 Q_1(r_1) r_1^2}{c^2}\right]\left[1.215 - \sqrt{1 + \frac{2\varepsilon^2 H_0^2 Q_1(r_1) r_1^2}{c^2}} + arctg\sqrt{1 + \frac{2\varepsilon^2 H_0^2 Q_1(r_1) r_1^2}{c^2}}\right] - 1 \tag{6.66}$$

When $H_0 r_1 / c \ll 1$, we have $Z_d \ll 1$, $Q_1(r_1) \rightarrow 1$ and $\varepsilon \approx 1$ as well as

$$Z_g \approx \frac{H_0^2 r_1^2}{2c^2} \approx \frac{Z_d^2}{2} \ll Z_d \tag{6.67}$$

It indicates that under the condition of low redshift, gravitational redshift can be neglected comparing with he Doppler redshift. But for the situations of high redshift just as supernova, gravitational redshift should be considered simultaneously.

Now let's discuss the cosmological redshift while the Doppler effect and the gravitational effect are considered simultaneously. Suppose that at moment $t_1$, the velocity of a celestial located at point $r_1$ is $V_1$. By the measurement of observers who move with a celestial body, the (proper) frequency and (proper) wave length of light emitted by the celestial body are $v_1$ and $\lambda_1$. But for the observers who are at lest on the static reference frame, light's frequency and wave length become $v_1'$ and $\lambda_1'$. That is to say, if there is a static observer who are near point $r_1$, he can only observe the Doppler redshift, instead of the gravitational redshift, at time $t_1$ when light has not began to move. Suppose that the frequency and wave length measured by the observer on the earth at last are $v_0$ and $\lambda_0$, the whole redshift in the process should be

$$Z = \frac{\lambda - \lambda_0}{\lambda_0} = \frac{\lambda - \lambda_1 + \lambda_1 - \lambda_0}{\lambda_0} = \frac{\lambda - \lambda_1}{\lambda_1}\frac{\lambda_1}{\lambda} + \frac{\lambda_1 - \lambda_0}{\lambda_0} \tag{6.68}$$

According to discussion above, the Doppler redshift and the gravitational redshift are individually

$$Z_d = \frac{\lambda - \lambda_1}{\lambda_1} \qquad\qquad Z_g = \frac{\lambda_1 - \lambda_0}{\lambda_0} \qquad (6.69)$$

So (6.68) can written as

$$Z = Z_d(Z_g + 1) + Z_g = Z_d + Z_g + Z_d Z_g \qquad (6.70)$$

This is just the cosmological redshift while the Doppler and the gravitational effects are considered simultaneously, in which the interference item between the Doppler redshift and gravitational redshift is contained. If there is no gravitation, the observers on the earth can only observe the Doppler redshift, no gravitational redshift. If there is the universal expansion, what can be observed is only gravitational redshift, no the Doppler redshift.

Meanwhile, we should reconsider the concept of the luminosity distance. As we know the luminosity distance $d_L$ is defined by the relation

$$B = \frac{L}{4\pi d_L^2} \qquad (6.71)$$

The absolute luminosity $L$ is determined by light's curve which is an absolute quantity, having nothing to do the practical distance of supernovae. But the brightness $B$ depends on practical distance between observer and illuminate celestial body. Suppose that at past time $t_1$, an illuminate celestial body located at point $r_1$ omitted $N$ photons during the period of time $\Delta t_1$. As discussed above, for the resting observers on the earth, owing to the Doppler effect, the frequency becomes $\nu$. Because photon has gravitational potential energy, photon's energy is $h\nu + U(r)$ in this case. So at this $t_1$ moment, the absolute luminosity of the celestial body is $L = N(h\nu + U)/\Delta t_1$. In this paper, for the static observers on the earth, the time is consistent in the universe, so these photons would arrive at the earth during the time $t_0$ and $t_0 + \Delta t_1$, and their frequencies would become $\nu_0$ for the observers on the earth with the brightness $B = Nh\nu_0 /(4\pi r_1^2 \Delta t_1)$. By considering the formula (4.105) of energy conservation, we have $h\nu + U(r) = h\nu_0$ and get $d_L = r_1$. That is to say, according to the theory of this paper, the luminosity distance is equal to practical distance. The definition of luminosity distance $d_L = r_1(1+Z)$ in the current cosmology is actually the production of violating energy conservation for photon's motion in gravitational field, we should abandon it.

As for the initial condition of the universal expansion and the Hubble constant, the theory is incapable. We should suppose the function form of $\kappa(b, r_0)$ and the value of $H_0$ in advance. Contrarily, if the form of $\kappa(b, r_0)$ has been known, we can understand the initial construction of the expansive universe. In practices, we can suppose the values of $H_0$ and $\kappa(b, r_0)$ for each supernova, then calculating its cosmological redshift. By comparing the result with practical observation, we can determine $\kappa(b, r_0)$ and $H_0$. After that, we can decide the function form of $\kappa(b, r_0)$ at last. But this is later work. In this paper, we only calculate several points to show that this theory is effective.

In Figure 6.2, we use the second curve from up to bottom represent the practically observed curve of Ia type high red-shift supernovae, corresponding to the values $\Omega_{0m} = 0.3$ and $\Omega_\lambda = 0.7$. In the figure, we have $m_B \approx 5.5 + 5\log d_L$ approximately. In Figure 6.3, the lowest curve can represent the practically observed result, in which we have $m_B \approx 5.5 + 5\log d_L$. By taking $H_0 = 65 Km \cdot Mpc^{-1}$ and the large distance $r_1 = 7 \times 10^{25} m$ at fist, we have $H_0 r_1 / c = 0.49$, $Q_1(r_1) = 0.612$ and $Q_2(r_1) = 0.785$. For the supernova under this condition, we take $\kappa = -0\cdot759$ and obtain $\varepsilon = 1.08$ from (6.75). In this case, we

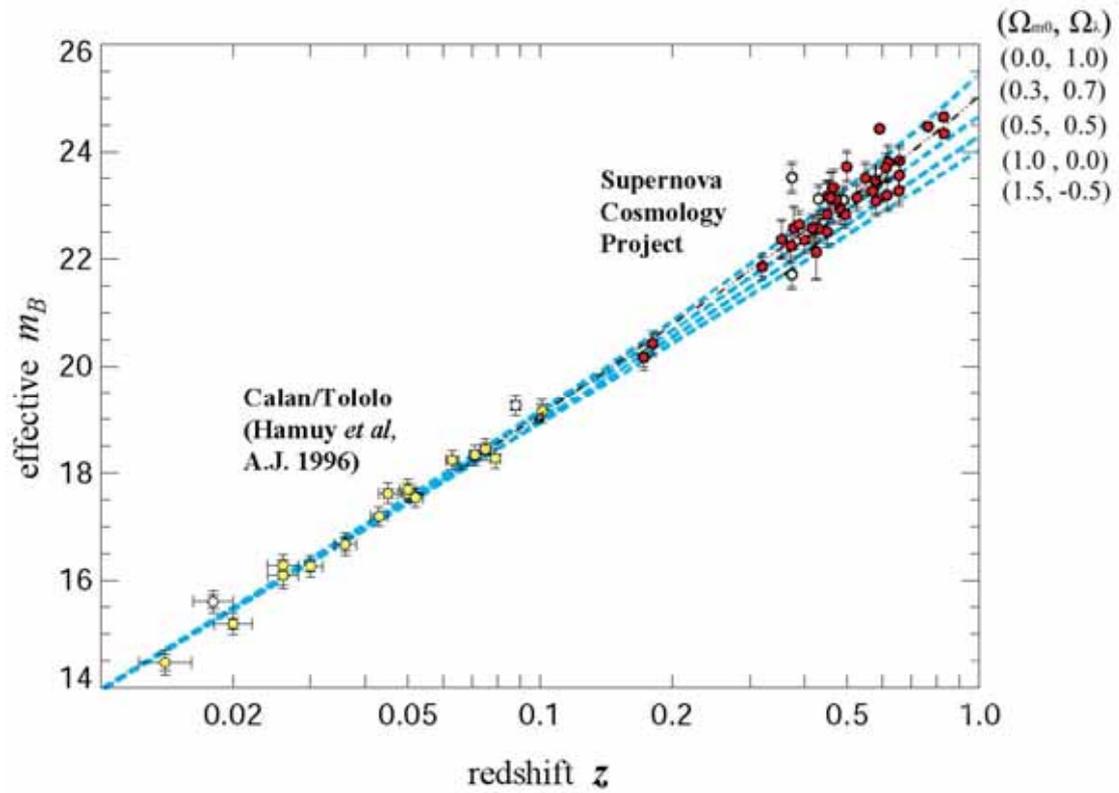

**Fig.6. 1 The Hubble diagram of Ia type high red-shift supernovae (1)**

**(The original diagram cited from Perlmutter S, et al, 1998)**

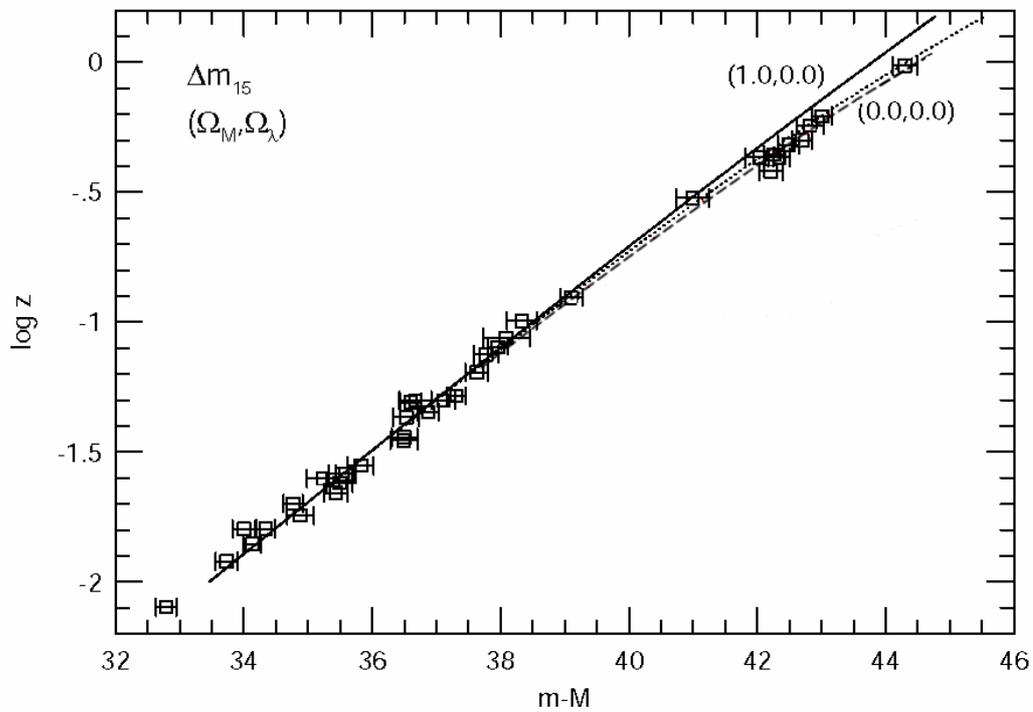

**Fig. 6.2 The Hubble diagram of Ia type high red-shift supernovae (2)**

**(The original diagram cited from Ricss A.G. et al, 1998)**

have $Z_g = 0.075$, and $Z_d = 0.275$ as well as $Z = 0.371$. Meanwhile, we have $m_B = 22.27$ and $m - M = 41.77$. This supernova is just located on the curve with $\Omega_{0m} = 0.3$ and $\Omega_\lambda = 0.7$. For the nearby celestial body with $r_1 \ll 10^{25} m$ or $\alpha / r_1 \ll 1$, taking $\kappa \approx 0$, we have $Z = H_0 r_1 / c$. The celestial body is also located on the straight line of redshift--- luminosity distance. For the supernova with $r_1 > 7 \times 10^{25} m$, the approximate formulas (6.47) and (6.48 are disabled, we need accurate method to calculate them.

Taking $H_0 = 32.5 Km \cdot Mpc^{-1}$, we calculate the cosmological redshift of celestial body at point $r_1 = 1.4 \times 10^{26} m$ and get $H_0 r_1 / c = 0.490$, $Q_1(r_1) = 0.612$ and $Q_2(r_1) = 0.785$. Let $\kappa = -0 \cdot 680$, according to (6.75), we have $\varepsilon = 1.25$. Therefore, we can get $Z_g = 0.092$ and $Z_d = 0.538$ as well as $Z = 0.680$. Meanwhile, we $m_B = 23.78$ and $m - M = 43.28$. This point is also approximately located on the curve corresponds to $\Omega_{0m} = 0.3$ and $\Omega_\lambda = 0.7$. For the nearby celestial body with $r_1 \ll 10^{26} m$ or $\alpha / r_1 \ll 1$, we take $\kappa = 3$ and get $Z \approx 2 H_0 r_1 / c$. They are located on the straight line of redshift--- luminosity distance. For the supernova with $r_1 > 1.4 \times 10^{26} m$, the approximate formulas (6.47) and (6.48 are also disabled, we need accurate method to calculate them.

So it is obvious that we can always get proper $\kappa(b, r_0)$ to calculate the cosmological redshift so that the results can coincide with practical observation. In this way, we can explain the cosmological redshift of supernova without introducing the hypothesis of dark energy and the accelerating expansion of the universe. In these two cases, when $r_1 \geq 7 \times 10^{25} m$, we always have $\kappa(b, r_0) < 0$, our universe may be a circulatory one to expanse and compress periodically.

## 6. The Hubble constant, the universal age and non-baryon dark material

According to (6.67), when $H_0 r_1 / c \ll 1$, we have $Z_g \ll Z_d$. So for nearby celestial bodies, their cosmological redshift are mainly the Doppler redshift, the gravitational redshft can be neglected. When $r_1 \ll 10^{26} m$, we have $Q_2(r_2) \to 1$. By considering (6.53), the Hubble experimental formula can be written as

$$Z_d = \frac{Hr_1}{c} = \frac{H_1 r_1}{c} \sqrt{Q_2(r_1) + \kappa(b, r_0)} = \frac{\varepsilon H_0 r_1}{c} \sqrt{Q_2(r_1) + \kappa(b, r_0)} \approx \frac{\varepsilon H_0 r_1}{c} \qquad (6.72)$$

In the formula, $H$ is the Hubble constant obtained by our practical measurements. It is obvious that $H$ is not a constant actually. $H_0$ is a constant determined by the practical material density of the universal. According to (6.62), we have

$$\frac{H_0 r_1}{c} = \frac{2(1 - 1/\varepsilon)}{3\sqrt{[Q_2(r_1) + Q_2(\varepsilon^{2/3} r_1)]/2 + \kappa}} \approx \frac{2}{3}\left(1 - \frac{1}{\varepsilon}\right) \qquad (6.73)$$

Let $H_0 r_1 / c = 10^{-3}$, we have $\varepsilon = 1.0015 \approx 1$. Taking $H_0 = 65 Km \cdot Mpc^{-1}$, we get $H \approx 65 Km \cdot Mpc^{-1}$. This is just the result of the current cosmology. If taking $H_0 = 32.5 Km \cdot s^{-1} \cdot Mpc^{-1}$, when $H_0 r_1 / c = 10^{-3}$, we still have $\varepsilon \approx 1$. By taking $\kappa = 3$, we get

$$Z_d = \frac{Hr_1}{c} = \frac{H_1 r_1}{c} \sqrt{Q_2(r_1) + \kappa(b, r_0)} \approx \frac{2 H_1 r_1}{c} = \frac{2 H_0 r_1}{c} \qquad (6.74)$$

We can introduce the concept of the equivalent material density $\rho_h$ of the universe. The position of $\rho_h$ is similar to the critical material density $\rho_c$ in the current cosmology. By using $\rho_h$, the Hubble experimental law can be written as

$$Z_d = \sqrt{\frac{8\pi G \rho_H}{3}} \frac{r_1}{c} = 2 \sqrt{\frac{8\pi G \rho_0}{3}} \frac{r}{c} = \sqrt{\frac{8\pi G 4 \times \rho_0}{3}} \frac{r_1}{c} \qquad (6.75)$$

So we have $\rho_0 = 0.25\rho_H$ from formula above. According to the observations of astronomy, we can take $H = 60 \sim 70 Km \cdot s^{-1} \cdot Mpc^{-1}$ or $\rho_h = 7.9 \times 10^{-27} kg/m^3$ for the current universe. According to the theory in this paper, we can take $H_0 = 30 \sim 35 Km \cdot s^{-1} \cdot Mpc^{-1}$ to explain the Hubble experimental law well. By taking medium value $H_0 = 32.5 Km \cdot s^{-1} \cdot Mpc^{-1}$, we have $\rho_0 = 1.98 \times 10^{-27} kg/m^3$. On the other hand, by means of photometry, the density of illuminant material in the universe is considered to be $\rho_0 \approx 2 \times 10^{-28} kg/m^3$ [23], corresponding to $\Omega_{m0} = \rho_0/\rho_h \approx 0.025$, at present day. By considering the fact that lots of non-illuminant material may exist within the galaxies and star clusters in the universe actually, it is rational to suppose that practical material density is about 10 times of illuminant material ($\Omega_{m0} \approx 0.25$). In this way, we have $\rho_0 \approx 2 \times 10^{-27} kg/m^3$. According to the synthesis theory of nucleus, the density of baryon material $\Omega_{b0} = \rho_b/\rho_h$ is determined by following formula [24]

$$\Omega_{b0} h^2 = 0.0037\zeta \qquad \zeta < 9 \qquad (6.76)$$

In light of the current theory, the Hubble constant corresponds to $h = 0.65$, so we have $\Omega_{b0} < 0.08$ from (17). Because of $\Omega_{m0} \approx 0.30$ from astronomical observations, it is thought at present that the synthesis theory of nucleus has proved that our universe is composed of non-baryon dark material mainly. But according to the new Hubble constant with $h = 0.325$, we have $\Omega_{b0} < 0.32$. It means that after the new Hubble constant is used, the universe should be composed of common baryon material mainly according to the synthesis theory of nucleus.

The practical Hubble constant may be between these two situations. So according to this paper, we can not need the hypotheses of non-baryon dark material again, but do not exclude its existence. The concept of non-baryon dark material has been put forward for decades, but up to now we can't really find them. In light of this paper, this problem may be eliminated. At least, we don't need to suppose more than three fourth of material in our universe to be non-baryon dark material, if they exist really.

The universal age can also be estimated by using (6.61). Taking $H_0 = 32.5 Km \cdot s^{-1} \cdot Mpc^{-1}$ and suppose that the celestial body moves from $r_1$ to $r_2 = 1.4 \times 10^{26} m >>$ with $r_1/r_2 << 1$. We have $Q_1(r_1) = 1$, $Q_2(r_2) = 0.785$ and $\kappa = -0.685$. Substitute them into (6.73), we obtain the universal age

$$\Delta t = \frac{2}{3 \times 0.456 H_0} = \frac{1.462}{H_0} = 4.386 \times 10^{10} \, y \qquad (6.77)$$

Taking $H_0 = 65 Km \cdot s^{-1} \cdot Mpc^{-1}$ and suppose that the celestial body moves from $r_1$ to $r_2 = 7 \times 10^{25} m$ with $r_1/r_2 << 1$, we also have $Q_1(r_1) = 1$, $Q_2(r_2) = 0.785$, $\kappa = -0.759$ and get the universal age

$$\Delta t = \frac{2}{3 \times 0.365 H_0} = \frac{1.826}{H_0} = 2.740 \times 10^{10} \, y \qquad 6.78$$

It is obvious that in theses two cases, there exists no the problem of the universal age again.

## 7. The motion equation of the universal expansion

Now let's discuss how to transform the formula (6.52) into the form of the current cosmology. Let $\bar{r}(t) = R(t)r$, $r$ is the coordinate of medium, $R(t)$ is the scalar factor of the expanding universe. Because we have not used the R-W metric, we have no the restriction of (6.2). There exists no the problem of violating the velocity addition ruler of special relativity. Let $t_0$ represent the current time, we have $\bar{r}(t_0) = r$, $R(t_0) = 1$. In this way, the formula (6.52) can be written as

$$\ddot{R} = -\frac{4\pi G}{3}\rho R\left(1 - \frac{6.13\pi G}{c^2}\rho R^2 r^2 + \frac{29.01\pi^2 G^2}{c^4}\rho^2 R^4 r^4\right.$$

$$\left.\cdots - \frac{125.53\pi^3 G^3}{c^6}\rho^3 R + \cdots\right) + \frac{1}{2}RA'(R)\delta(R-R_0) \quad (6.79)$$

Because of $\rho(t)R^3(t) = \rho(t_0)R^3(t_0) = $ constant, we can let

$$\frac{6.13\pi G}{c^2 R(t_0)}\rho(t)R^3(t)r^2 = \frac{6.13\pi G}{c^2}\rho(t_0)R^2(t_0)r^2 = b_1' \quad (6.80)$$

$$\frac{29.01\pi^2 G^2}{c^4 R^2(t_0)}\rho^2(t)R^6(t)r^4 = \frac{6.13\pi G}{c^2}\rho^2(t_0)R^4(t_0)r^4 = b_2' \quad (6.81)$$

$$\frac{125.53\pi^3 G^3}{c^4 R^3(t_0)}\rho^3(t)R^9(t)r^6 = \frac{6.13\pi G}{c^2}\rho^3(t_0)R^6(t_0)r^6 = b_3' \quad (6.82)$$

Here $b_i'$ are dimensionless parameters represented by the current time. Let $R(t_0) = R_0$, the formula (6.79) can be written as again

$$\frac{\ddot{R}}{R} = -\frac{4\pi G}{3}\rho\left(1 - \frac{b_1' R_0}{R} + \frac{b_2' R_0^2}{R^2} - \frac{b_3' R_0^3}{R^3} + \cdots\right) + \frac{1}{2}A'(R)\delta(R-R_0) \quad (6.83)$$

By taking the integral, we obtain

$$\left(\frac{\dot{R}}{R}\right)^2 + \frac{K(R_0)}{R^2} = \frac{8\pi G}{3}\rho\left(1 - \frac{b_1}{R} + \frac{b_2}{R^2} - \frac{b_3}{R^3}\cdots\right) \quad (6.84)$$

In Which $b_1 = b_1'R_0/2$, $b_2 = b_2'R_0^2/3$, $b_3 = b_3'R_0^3/4\cdots$. Comparing with the standard form in the current cosmological theory, the item relative to the universal constant is replaced by the items containing $b_i$. In the current cosmology, $K$ is a curvature constant. But in (6.84), $K$ is an integral constant. Let

$$\rho_1' = -\rho(-b_1'R_0/R + b_2 R_0^2/R^2 - \cdots)/2 \qquad \rho_2' = \rho(-b_1/R + b_2/R^2 - \cdots) \quad (6.85)$$

We can re-write (6.83) and (6.84) as

$$\frac{\ddot{R}}{R} = -\frac{4\pi G}{3}(\rho - 2\rho_1') + \frac{1}{2}A'(R)\delta(R-R_0) \quad (8.86)$$

$$\left(\frac{\dot{R}}{R}\right)^2 + \frac{K(R_0)}{R^2} = \frac{8\pi G}{3}(\rho + \rho_2') \quad (6.87)$$

Comparing with the current cosmology, besides there is more additional item in (6.86), the difference is on the definitions $\rho_\lambda, \rho_1', \kappa$ and $K(R_0)$. In the current theory, $\rho_\lambda$ is the effective energy density corresponding to vacuum and the cosmic constant. In order to coincide with the observation of high redshift Ia supernova, we have to suppose $\rho - 2\rho_\lambda < 0$ for the current universe, so that the concept of dark energy has to be introduced and we have to think that the universal expansion is being accelerated at present. It is difficult for us to imagine what dark energy is. How can we find it? According to this paper, we always have $\rho - 2\rho_1' > 0$. There is no the problem of the universal accelerating expansion. In fact, there exists a strident contradiction about cosmic constant. This problem puzzled physicists for a long time can be cast off thoroughly in light of this paper.

It is noted that the discussion of this paper is based on the Schwarzschild solution of the Einstein's equation of gravitational field. What we do is to transform its geodesic equation into flat space-time to describe. In this way, we can rationally establish the similar equation of cosmology and describe the problem of cosmology well. Because the theory is based on flat space-time, we have no the problem of the universal flatness again. Because the magnetic-like gravitation exists, it may be used to solve the problem of galaxy formation and so do. Based on this foundation, we would establish a really rational theory of cosmology without any singularity.

## 8. The possibility to explain the Pioneer Anomaly

NASA had lunched a series of spacecrafts such as the Pioneers 10 and 11 for the exploration of `the Jupiter and Saturn since 1970's last century. Recently, it is founded that most of the spacecrafts departure their orbits. The orbits are calculated by the Einstein's theory of gravitation. An additional constant acceleration is found for them with $a_p = (8.74 \pm 1.33) \times 10^{-10} m/s^2$ according to the data up to date [25]. The direction of $a_p$ is towards the sun. After all of possible factors, which would affect the orbits of the spacecrafts, were excluded, scientists in NASA affirmed that there exists the unexplained Pioneer Anomaly.

Since the Pioneer Anomaly was founded in 1998, many theories had been proposed, but none of them was satisfied [26]. The orbits of the spacecrafts and the acceleration $a_p$ are calculated actually by the PPN approximate method based on the Einstein's theory. So the Pioneer Anomaly means that the Einstein's theory of gravitation would be revised. New we discuss the possibility to explain the Pioneer Anomaly in light of this paper's theory. The accelerations of spacecrafts are calculated by using the formula below [27]

$$\vec{a}_i = \sum_{j \neq i} \frac{\mu_i (\vec{r}_j - \vec{r}_i)}{r_{ij}^3} \left\{ 1 - \sum_{k \neq i} \frac{4\mu_k}{c^2 r_{ik}} - \sum_{k \neq j} \frac{\mu_k}{c^2 r_{jk}} - \frac{3[(\vec{r}_j - \vec{r}_i) \cdot \vec{V}_j]^2}{2c^2 r_{ij}^2} + \frac{(\vec{r}_j - \vec{r}_i) \cdot \vec{a}_j}{2c^2} \right.$$

$$\left. - \frac{4\vec{V}_i \cdot \vec{V}_j}{c^2} + \frac{\vec{V}_i^2}{c^2} + \frac{2\vec{V}_j^2}{c^2} \right\} + \sum_{j \neq i} \frac{\mu_j (\vec{V}_i - \vec{V}_j)}{c^2 r_{ij}^3} [(\vec{r}_j - \vec{r}_i) \cdot (4\vec{V}_i - 3\vec{V}_j)] + \sum_{j \neq i} \frac{7\mu_j \vec{a}_j}{2c^2 r_{ij}} \qquad (6.88)$$

Here $r_{ij} = |\vec{r}_j - \vec{r}_i|$, $\mu_i = GM_i$. $M_i$, $\vec{a}_i$ and $\vec{V}_i$ are the $i$ object's static mass acceleration and velocity individually. If only a spacecraft moves in the static gravitational field of the sun, we can let $\vec{V}_2 = \vec{a}_2 = 0$, $\vec{r} = \vec{r}_{12}$, $\vec{V} = \vec{V}_1$ and get spacecraft's acceleration from the formula above

$$\vec{a} = -\frac{GM\vec{r}}{r^3} \left( 1 - \frac{4GM}{c^2 r} - \frac{4Gm}{c^2 r} + \frac{\vec{V}^2}{c^2} \right) + \frac{4GM(\vec{r} \cdot \vec{V}) \vec{V}}{c^2 r^3} \qquad (6.89)$$

Because the PPN approximate method is also based on curved space-time actually, the acceleration shown in Eq.(6.88) can not be compared with the experiments carried on the earth before it is transformed into the result in flat space-time. So it is meaningless actually. According to the paper, we should calculate gravitational interaction among the sun, planets and spacecrafts based on Eq.(5.20). Here we only use Eq.(5.2) to show the acceleration of spacecraft. It should be

$$\vec{a} = \frac{d^2\vec{r}}{dt^2} = -\frac{GM\vec{r}}{r^3} \left( 1 + \frac{3L^2}{c^2 r^2} \right) \left( 1 - \frac{V^2}{c^2} \right) - \frac{\dot{V} V \vec{V}}{c^2 (1 - V^2/c^2)} \qquad (6.90)$$

In which $V$ and $\dot{V} = dV/dt$ are determined by Eq.(4.22). It can be seen that the formula (6.90) is different from Eq.(6.89). The formula (6.90) is an accurate result relative to angle momentum. But Eq.(6.89)

is an approximate result having nothing to do with angle momentum. For simplification, we only discuss the first item of Eq.(6.90). When $\alpha \ll r$, $V \ll c$, by remaining the items up to the order $r^{-4}$, we have

$$a = -\left(1 + \frac{3L^2}{c^2 r^2} - \frac{GM}{c^2 r}\right)\frac{GM}{r^2} \qquad \Delta a = -\left(\frac{3L^2}{c^2 r^2} - \frac{GM}{c^2 r}\right)\frac{GM}{r^2} \qquad (6.91)$$

Here $\Delta a$ is just the additional acceleration comparing with the Newtonian theory. Suppose that the spacecraft moves along the tangent direction near the solar surface in the third universal velocity $V = 4.20 \times 10^4 M/s$, so that it can escape the sun's gravitation. The sun's mass is $M = 1.99 \times 10^{30} Kg$, the angle momentum of spacecraft is $L = VR$. The additional acceleration is a positive value with $\Delta a = 5.64 \times 10^{-4} m/s^2$. The result means that the revised force is a repulsive one, instead of gravitation. When the spacecraft is on the mercurial orbit, we get $\Delta a = -1.38 \times 10^{-9} m/s^2$. If the spacecraft is on the earth's orbit, we have $\Delta a = -2.93 \times 10^{-10} m/s^2$ with the same magnitude of the Pioneer anomaly $a_p$. When spacecraft is on the Jupiter's orbit, we have $\Delta a = -6.38 \times 10^{-14} m/s^2$. It becomes very small. When spacecraft moves on the earth's surface in the second universal velocity $V = 1.12 \times 10^4 m/s$, we have $\Delta a = -3.41 \times 10^{-8} m/s^2$. So the earth, Jupiter and other planets would also cause additional accelerations for spacecraft. Especially, the Jupiter and Saturn's masses are quite big, the additional accelerations should be taken into account.

In the practical calculations, the magnetic-like gravitation caused by the motions of planets should also be considered. The time delay effects of radar waves emitted by spacecraft should also be calculated by means of Eq.(4.74). At last, the really strict calculations should be carried out in the absolutely reference frame. Because the multi-body problems are involved, the practical orbits of spacecrafts should be calculated by computer. By comparing the results of numerical value calculations with the practical orbits of spacecraft, we can judge whether or not the theory of this papers is more rational than the general relativity. If it is alight, we would reach a really rational theory of gravitation, and obtain a credible foundation for the unification of four interaction forces.

To sum, in order to explain the light's propagations in vacuum a hundred years ago, the hypothesis of the ether with very strange natures was putted forward. In order to eliminate the ether, Einstein established special relativity. After that, general relativity was advanced. The theories caused the idea revolution of space-time and gravitation and promoted the development of physics. A hundred year later, in order to explain so many contradictions and anomalies, we foist too many things such as cosmic constant, vacuum energy, dark energy and space-time singularity and so on into vacuum again. The situation actually indicates that we need another idea renewal on space-time and gravitation. The result would be that these concepts would be given up at last, just as that the concept of the ether was given up a hundred years ago.